\documentclass[structabstract]{aa}  
\usepackage{graphicx}
\usepackage{txfonts}
\usepackage{natbib}
  \def\ProDiMo{{\sc ProDiMo\ }}

  \def\cT2{c_T^2}

  \def\eg{e.\,g.\ }

  \def\amin{{a_{\rm min}}}
  \def\amax{{a_{\rm max}}}

  \def\mcfost{{\sf MCFOST}}
\def\ProDiMo{{\sf ProDiMo}}

\begin{document}

%   \title{Disk structure and evolution for Herschel}
   \title{Continuum and line modeling of disks around young stars}

   \subtitle{II. Line diagnostics for {\sf GASPS} from the {\sf DENT} grid}

   \author{I. Kamp\inst{1}
         \and
          P. Woitke\inst{1,2,3,4}
          \and
          C. Pinte\inst{3}
         \and
          I. Tilling\inst{5}
          \and
          W.-F. Thi\inst{3}
          \and
          F. Menard\inst{3}
          \and
          G. Duchene\inst{6,3}
          \and
         J.-C. Augereau\inst{3}
            }

         \institute{
             Kapteyn Astronomical Institute, Postbus 800,
             9700 AV Groningen, The Netherlands
         \and
             UK Astronomy Technology Centre, Royal Observatory, Edinburgh,
             Blackford Hill, Edinburgh EH9 3HJ, UK
        \and
             UJF-Grenoble 1 / CNRS-INSU, Institut de Plan\'etologie et d'Astrophysique de Grenoble (IPAG) UMR 5274, 38041 Grenoble Cedex 9, France
          \and
             School of Physics \& Astronomy, University of St.\ Andrews, North Haugh, St. Andrews KY16 9SS, UK
         \and
	     SUPA, Institute for Astronomy, Royal Observatory, Edinburgh,
             Blackford Hill, Edinburgh EH9 3HJ, UK
          \and
             Astronomy Department, University of California, Berkeley, CA 94720-3411, USA
            }

   \date{Received ; accepted }

% \abstract{}{}{}{}{} 
% 5 {} token are mandatory
 
\abstract
   {}
   % aims (mandatory)
    {We want to understand the chemistry and physics of disks on the basis of a large unbiased and statistically relevant grid of disk models. One of the main goals is to explore the diagnostic power of various gas emission lines and line ratios for deriving main disk parameters such as the gas mass.}
   % methods heading (mandatory)
    {We explore the results of the {{\sf DENT}} grid (\underline{\bf D}isk \underline{\bf E}volution with \underline{\bf N}eat \underline{\bf T}heory) that consists of 300\,000 disk models with 11 free parameters. Through a statistical analysis, we search for correlations and trends in an effort to find tools for disk diagnostic. }
   % results (mandatory)
    {All calculated quantities like species masses, temperatures, continuum and line fluxes differ by several orders of magnitude across the entire parameter space. The broad distribution of these quantities as a function of input parameters shows the limitation of using a prototype T Tauri or Herbig Ae/Be disk model. The statistical analysis of the {{\sf DENT}} grid shows that CO gas is rarely the dominant carbon reservoir in disks. Models with large inner radii (10 times the dust condensation radius) and/or shallow surface density gradients lack massive gas phase water reservoirs. Also, 60\% of the disks have gas temperatures averaged over the oxygen mass in the range between 15 and 70~K; the average gas temperatures for CO and O differ by less than a factor two. Studying the observational diagnostics, the [C\,{\sc ii}]\,158~$\mu$m fine structure line flux is very sensitive to the stellar UV flux and presence of a UV excess and it traces the outer disk radius ($R_{\rm out}$). In the submm, the CO low J rotational lines also trace $R_{\rm out}$. Low [O\,{\sc i}]\,63/145 line ratios ($<$ a few) can be explained with cool atomic O gas in the uppermost surface layers leading to self-absorption in the $63~\mu$m line; this occurs mostly for massive non-flaring, settled disk models without UV excess. A combination of the [O\,{\sc i}]\,63 line and low J CO lines correlates with several disk properties such as the average O\,{\sc i} gas temperature in disks, the outer disk radius, and the UV excess. }
   % conclusions (optional)
    {The [O\,{\sc i}]\,63/CO~2-1 line ratio is a powerful diagnostic to break disk modeling degeneracies. A combination of the [O\,{\sc i}]\,63~$\mu$m flux and the [O\,{\sc i}]\,63/CO~2-1 line ratio can be used for $M_{\rm gas} \leq 10^{-3}$~M$_\odot$ to obtain an order of magnitude estimate for the disk gas mass purely from gas observations. The previously used conversion of a CO submm line flux alone generally leads to larger uncertainties.}



   \maketitle

%=====================================================================

\section{Introduction}

In the recent decade, the modeling of the gas chemistry and physics in protoplanetary disks has become a field of its own, following the lead of the dust SED modeling with multi-dimensional continuum radiative transfer {\citep[e.g.][]{Chiang1997,Dalessio1998, Dullemond2002}.} The impact of the gas chemistry and energy balance is largest in the disk surface layers where most of the infrared and sub-mm line emission is arising \citep[e.g.][]{Kamp2004,Jonkheid2004,Gorti2004,Nomura2005}. The complexity of the gas chemistry and individual heating/cooling processes as well as their mutual dependence makes the gas structure modeling more complicated, physically more demanding, and computationally more time consuming. Hence, most modeling has so far concentrated on either single selected objects \citep[e.g.][]{Qi2003,Semenov2005,Qi2008,Gorti2008,Henning2010}, prototypes \citep[e.g.][]{Aikawa2002,Meijerink2008,Goicoechea2009} or small model series studying the dependencies of specific parameters on the gas emission from disks \citep[e.g.][]{Aikawa2006,Jonkheid2007,Nomura2007,Glassgold2009,Kamp2010}.

%With the advent of many upcoming observing facilities that will study the gas in protoplanetary disks such as Herschel, ALMA, JWST/MIRI, SPICA/SAFARI etc., there is a clear need for a simultaneous and consistent modelling of dust and gas in protoplanetary disks, with the aim to identify the diagnostic power of certain continuum and line observations for the physical, chemical and temperature structure in the disks. 
Current and upcoming observing facilities such as Herschel, ALMA, JWST/MIRI, SPICA/SAFARI are excellent tools to study the gas in large samples of protoplanetary disks. Hence, there is a clear need for a simultaneous and consistent modelling of dust and gas in protoplanetary disks, with the aim to identify the diagnostic power of certain continuum and line observations for the physical, chemical and temperature structure in the disks. 
These identifications must be based on a large variety of disk models covering a wide spread in stellar, disk, dust, and gas parameters.

%The theory groups in Edinburgh, Grenoble and Groningen  
We jointly produced a grid of $300\,000$ disk models ((\underline{D}isc \underline{E}volution with \underline{N}eat \underline{T}heory) to study the predictive power of individual gas lines, line ratios and continuum tracers \citep{Woitke2010}. The line emission studies concentrate on a series of fine structure (O\,{\sc i} and C\,{\sc ii}) and molecular lines (high J CO and H$_2$O) that are observable with Herschel as well as a number of CO sub-mm lines accessible from the ground. A first analysis of the grid has shown that for example the [O\,{\sc i}]\,63\,$\mu$m emission strongly depends on the amount of UV excess coming from the star \citep{Woitke2010}. Subsequent application to the first findings from the Herschel Open Time Key Program {\sc GASPS} \citep{Pinte2010} revealed that the measured [O\,{\sc i}] fluxes for a sample of ~30 disks indicate that disks around T Tauri stars indeed require extra heating of the disk surface --- either through a UV excess and/or X-ray irradiation ---, while the emission from disks around Herbig Ae stars can generally be matched by photospheric heating alone. This paper now presents a more systematic and complete study of the gas properties and diagnostics from the {\sf DENT} grid. The dust diagnostics will be discussed in a separate paper.

In the following, we briefly summarize the {\sf DENT} grid approach (Section~\ref{DENT}) as presented in \citet{Woitke2010} before we discuss the statistics of gas chemistry and thermal structure of the $300\,000$ models in Sect.~\ref{chemistry} and the line diagnostics in Sect.~\ref{diagnostics}. We present our current best strategies for deriving disk gas parameters in the final conclusions (Sect.~\ref{conclusions}).

\section{The {\sf DENT} grid}
\label{DENT}

The {\sf DENT} grid has been described for the first time in \citet{Woitke2010} and we summarize here only a few key aspects. More details can be found in the appendix. 

The grid is computed using the 3D Monte Carlo radiative transfer code \mcfost\ \citep{Pinte2006,Pinte2009} and the gas chemistry and physics modules of \ProDiMo\ \citep{Woitke2009a,Kamp2010}. The column densities, flaring, dust size distribution and settling are parametrized with power laws, assuming a prescribed global gas-to-dust ratio. Mie theory is used to compute the dust optical properties assuming an astronomical silicate composition. The central star is parametrized by its mass and age, as well as strength of the UV excess. Dust and gas temperatures are calculated separately from Monte Carlo continuum radiative transfer and a detailed heating/cooling network, respectively (see Sect.~\ref{MCFOST} and \ref{ProDiMo}). 

We have chosen 11 free model parameters for the grid, leading to a total of $322\,030$ disk models. The range of parameter values explored in {\sf DENT} is discussed further in the appendix. The total computing time of the grid is 200\,000~CPU hours, an average of 40~CPU minutes per model. The total grid took 3 weeks on 400~processors and generated 1.8~TB of data.

The grid is very coarse in nature, e.g.\ stepping through gas mass in factors of ten. Hence, we cannot expect any method that we develop to measure disk properties such as the disk gas mass from gas line emission or line ratios to be more accurate than that step size, in this case an order of magnitude. The same holds for other quantities such as the outer disk radius, the flaring index, the dust-to-gas mass ratio etc..

For the following analysis, it is also important to keep in mind that the grid of disk models is not constructed to reflect the statistics with which disks occur in nature, e.g.\ the fact that some objects are rarer than others. There could be some disk types that we missed in our approach, or others that simply do not occur in nature. We take the latter into account by excluding certain parameter combinations in the analysis of the grid. Models with a flaring index of $\beta=0.8$ are for example more appropriate for late stages of disk evolution such as debris disks and so they are excluded from the study of line diagnostics in younger disks (noted in the figure captions whenever appropriate). A global disk analysis using the Toomre criterium shows that 94\% of the disk models in the {\sc Dent} grid are gravitationally stable (see Appendix~\ref{Stability}).

We chose parametrized density structures to avoid imposing the complex gas structure that is based on presumptions mainly derived from disk modeling approaches such as those of \citet{Gorti2004},  \citet{Nomura2005} or \citet{Woitke2009a}. While it is true that the more consistent vertical disk structure models need less assumptions than the simple parametrized models used in {\sf DENT}, even a full \ProDiMo\ including the vertical structure calculation \citep{Woitke2009a} is not guaranteed to capture all relevant physics. Using parametrized disk structures enables us to explore a large unbiased parameter space. From these simplified disk structures, we can still assess the relative influence of several of the key parameters on line and continuum fluxes. Even though the modeled fluxes are not absolute, we can still test fundamental aspects of our disk understanding, such as uncertainties in various observational techniques to derive disk dust and gas masses.

\subsection{\mcfost}
\label{MCFOST}

\mcfost\ is 3D Monte Carlo continuum and line radiative transfer code. The
details of the numerical schemes are presented in \cite{Pinte2006} and
\cite{Pinte2009}. In short, the code stochastically propagates 
photon packets in 3D through the disk. The transport of packets is
governed by successive scattering, absorption and re-emission
events, which are determined by the local dust properties (opacity, albedo,
scattering matrix) and temperature. \mcfost\ uses the immediate re-emission concept
\citep{Bjorkman2001} with a continuous deposition of energy to estimate
the mean intensity \citep{Lucy1999}. In the optically thick regions of
the disk, the code uses a diffusion approximation method to converge
the temperature structure. Emerging SEDs are generated by a
ray-tracing run from the temperature and radiation field estimated by
the Monte Carlo run.

For the {\sf DENT} grid, $10^6$ packets were used to compute the temperature
structure. The radiation field was converged using $10^4$ packets per wavelength
bin at wavelength larger than 0.5\,$\mu$m. As a reliable estimate of
the UV field is critical for the chemistry module in \ProDiMo, we
used $10^5$ packets per wavelength bin for wavelengths shorter than
0.5\,$\mu$m. This ensures  that the UV radiation field is well
converged, even in the deep regions of the disk.

\mcfost\ also includes a NLTE line transfer module which is not fully used
here, but whose ray-tracing module is used instead to compute the
emerging spectral lines from the level populations estimated by \ProDiMo.

\subsection{\ProDiMo}
\label{ProDiMo}

\ProDiMo\ calculates radiation thermo-chemical models of protoplanetary disks. It is described in detail in \citet{Woitke2009a} and \citet{Kamp2010} and we summarize here the most salient features. Most noticable, the code is not used in its full capacity, because the self-consistent computation of the vertical hydrostatic equilibrium structure is omitted. 
%This is discussed in some more detail in Sect.~\ref{DENT}.

The code takes as input the dust density and temperature, $n_{\rm dust}(r,z)$, $T_{\rm dust}(r,z)$ and dust properties as provided by \mcfost. In addition, the mean radiation field $J_\nu(r,z)$ from \mcfost\ is passed on to \ProDiMo. There are two grid parameters that apply only to the gas part of the model: the gas mass $M_{\rm gas}$, and hence the dust-to-gas mass ratio $\delta$, and the fractional UV luminosity defined as $f_{\rm UV} = L_{\rm UV}/L_\ast$ in the wavelength range between 91 and 250~nm.

The chemical network consists of 9 elements, 71 species connected through 950 reactions (neutral-neutral, ion-molecule, photoreactions, cosmic ray reactions and adsorption \& desorption of CO, CO$_2$, H$_2$O, NH$_3$, CH$_4$). The background radiation field $J_\nu$ enters the computation of the gas chemical and energy balance at several points, namely in the detailed frequency dependent integration of the UV ionization and dissociation cross sections, in the radiative pumping rates involved in the non-LTE modeling of atoms and molecules (O\,{\sc i}, C\,{\sc i}, C\,{\sc ii}, Mg\,{\sc ii}, Fe\,{\sc ii}, Si\,{\sc ii}, S\,{\sc ii}, CO rotational \& ro-vibrational, o-H$_2$ \& p-H$_2$ ro-vibrational, o-H$_2$O \& p-H$_2$O rotational), and in the photoelectric heating rates. The atomic and molecular data used in the statistical equilibrium calculation is described in detail in Sect.~6.1.5 of \citet{Woitke2009a}.

%From a few case studies, where we ran a full \ProDiMo model for the same parameter combination, we find that the lines are generally stronger than those found from the {\sc DENT} grid. 
%%However, this is not saying that one or the other approach is more correct. 
%At this stage, we are only starting to evaluate the available far-IR (Herschel) and sub-mm lines for larger samples of objects and develop a better understanding of the processes that determine the vertical disk structure. \remark{There might be more to add here after all; I need to revisit old notes and emails.}

\section{Gas physics and chemistry}
\label{chemistry}

As a first step in the statistical analysis of the {\sf DENT} grid, we discuss here how the masses and average temperatures of a number of key species (O, CO, C$^+$ and H$_2$O) depend on the grid parameters (see Appendix~\ref{Methodology}). With this method, we can obtain a first understanding of the disk chemistry (atomic versus molecular), the relevance of ices (how much CO or H$_2$O is bound in ice form) and the thermal disk structure (gas dust de-coupling and cool surface layers).

\subsection{Oxygen and CO masses}

Oxygen is the third most abundant element in the interstellar medium. In dense molecular clouds or protoplanetary disks, oxygen can be either in the gas or ice phase (both are included in our chemical model); in the gas phase, it can be atomic (O) or molecular (CO, O$_2$, H$_2$O). By measuring all possible phases, we can constrain the total oxygen mass in a disk. 
%Using a typical oxygen abundance, this can then be converted into total gas masses.

\begin{table}[h]
\caption{Maximum possible fraction of total disk mass for selected species $X_{\rm sp}$ assuming all carbon or oxygen are locked into that species. Also given are the species mass ($m_{\rm sp}$), abundance ($\epsilon$) and the expected values from the statistical distribution of gas mass fractions.}
\begin{tabular}{l | lllll}
\hline
species & $m_{\rm sp}$  & $\epsilon$ & $X_{\rm sp}$ & $\langle X_{\rm sp} \rangle$  & $\sigma$ \\
               & [m$_{\rm H}$] & w.r.t. H         & maximum  & EV &  \\ 
\hline\\[-2mm]
O            & 16         & 2.90 (-4)  & 3.5 (-3) & 2.3 (-3) & 1.2(-3) \\[1mm]
               & \multicolumn{3}{l}{only $10^{-5}<M_{\rm disk}\leq 10^{-6}$~M$_\odot$} & 3.4 (-3) & 2.2 (-4) \\
               & \multicolumn{3}{l}{only $10^{-1}<M_{\rm disk}\leq 10^{-2}$~M$_\odot$} & 1.2 (-3) & 7.1 (-4) \\
               \hline\\[-2mm]
C$^+$   & 12         & 1.30 (-4)  & 1.2 (-3) & 4.5 (-4) & 5.0(-4) \\[1mm]
               & \multicolumn{3}{l}{only $10^{-5}<M_{\rm disk}\leq 10^{-6}$~M$_\odot$} & 1.0 (-3) & 2.5 (-4) \\
               & \multicolumn{3}{l}{only $10^{-1}< M_{\rm disk}\leq 10^{-2}$~M$_\odot$} & 9.4 (-6) & 2.3 (-5) \\
               \hline\\[-2mm]
CO         &  28        & 1.30 (-4)  & 2.8 (-3) & 7.9 (-4) & 8.5(-4) \\[1mm]
               & \multicolumn{3}{l}{only $10^{-5}<M_{\rm disk}\leq 10^{-6}$~M$_\odot$} & 1.4 (-3) & 2.9 (-3) \\
               & \multicolumn{3}{l}{only $10^{-1}< M_{\rm disk}\leq 10^{-2}$~M$_\odot$} & 1.3 (-3) & 8.9 (-4) \\
\end{tabular}
\end{table}

%\begin{figure}
%\centering
%\includegraphics[width=9cm]{MOMCO.eps}
%\caption{Left panel: The atomic oxygen mass fraction $M({\rm O})/M_{\rm gas}$ for all disk models. Color coded is the total disk gas mass. Right panel: The CO mass fraction $M({\rm CO})/M_{\rm gas}$ for disk models with $M_{\rm gas} \geq 10^{-5}$~M$_\odot$. Color coded is the dust-to-gas mass ratio.}
%\label{fig:MOMCO}
%\end{figure}

\begin{figure}
\centering
\includegraphics[width=9cm]{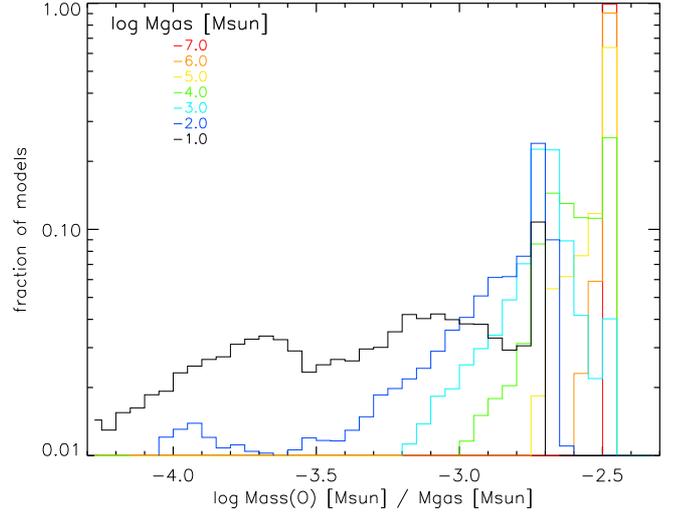}
\caption{The atomic oxygen mass fraction $X({\rm O}) = M({\rm O})/M_{\rm gas}$ for all disk models with $M_{\rm gas} \geq 10^{-7}$~M$_\odot$. Color coded is the total disk gas mass. Note that the y-axis is logarithmic.}
\label{fig:histMO}
\end{figure}

\begin{figure}
\centering
\includegraphics[width=9cm]{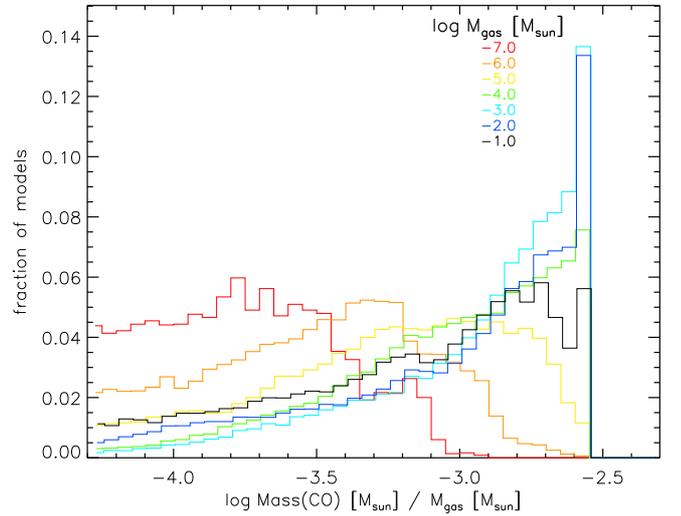}
\caption{The CO gas mass fraction $X({\rm CO}) = M({\rm CO})/M_{\rm gas}$ for all disk models with $M_{\rm gas} \geq 10^{-7}$~M$_\odot$. Color coded is the total disk gas mass.}
\label{fig:histMCO_Mgas}
\end{figure}

\begin{figure}
\centering
\includegraphics[width=9cm]{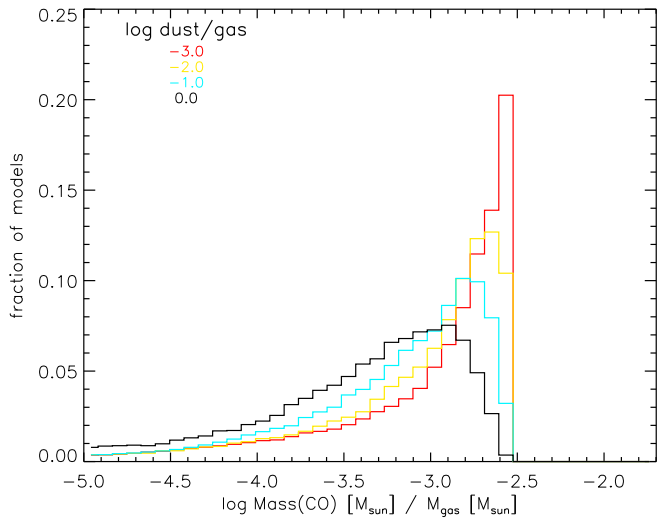}
\caption{The CO gas mass fraction $X({\rm CO}) = M({\rm CO})/M_{\rm gas}$ for all disk models. Color coded is the dust-to-gas mass ratio.}
\label{fig:histMCO}
\end{figure}

Fig.~\ref{fig:histMO} shows the statistical distribution of atomic oxygen masses. Low mass disk models have almost no CO, H$_2$O or ices ($X({\rm CO}) \lesssim 10^{-3}$, $X({\rm H_2O}) \lesssim10^{-4}$) and hence their atomic oxygen mass fraction is close to maximum. Assuming all oxygen is atomic, we obtain a mass fraction of $X({\rm O})=2.3 \times 10^{-3}$. Higher mass disk models show a wider distribution of atomic oxygen mass fractions in Fig.~\ref{fig:histMO}. This is due to a higher molecular fraction (dust shielding and gas self-shielding), e.g.\ CO, CH$_4$, and H$_2$O. Studying the CO mass fraction distribution (Fig.~\ref{fig:histMCO_Mgas}), the number of disks with masses larger than $10^{-5}$~M$_\odot$ that have a particular CO mass fraction is rising all the way towards the maximum possible CO mass fraction. The lowest mass disks ($10^{-6} \leq M_{\rm gas} \leq 10^{-7}$~M$_\odot$) show instead a broader and flatter distribution; ice formation and photodissociation for those models are extremely model dependent. Also, none of those models ever shows all carbon being bound in CO. The dust-to-gas mass ratio lowers for example the peak of the CO mass fraction by a factor 5 going from $0.001$ to $1$ (Fig.~\ref{fig:histMCO}). This is due to the average dust temperature being lower in the high dust-to-gas models and hence more CO freezing out.

\subsection{Water masses}

\begin{figure}
\centering
\includegraphics[width=9cm]{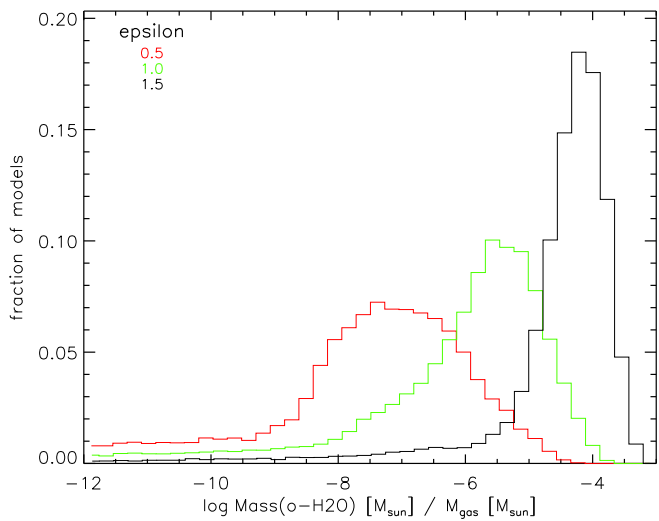}
\caption{The H$_2$O mass fraction $X({\rm H_2O}) = M({\rm H_2O})/M_{\rm gas}$ for disk models with an inner radius equal the sublimation radius. Color coded is the surface density power law index $\epsilon$.}
\label{fig:histMH2O}
\end{figure}

The mass fraction of gas phase water depends strongly on the surface density power law index, i.e.\ on the distribution of mass within the disk. Massive disk models that concentrate the mass in the inner dense disk ($\epsilon = 1.5$), convert a substantial fraction of their oxygen into water (Fig.~\ref{fig:histMH2O}). For shallower density gradients, the outer disk contains more mass than the inner disk; here temperatures are low so that water is predominantly frozen on the grain surfaces. What remains is a low fraction of gas phase water above the ice layer. Models that have an inner gap, i.e.\ $R_{\rm in}=10$ or 100 times the sublimation radius tend to have lower mass fractions of gas phase water. Those models lack the big inner gas phase water reservoir \citep{Woitke2009b}.

\subsection{Oxygen gas temperatures}
\label{Sect:OITg}

\begin{figure}
\centering
\includegraphics[width=9cm]{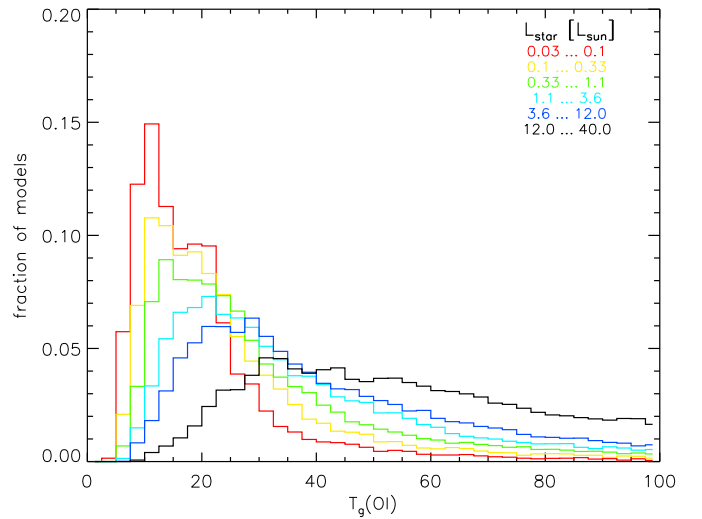}
\caption{Mean gas temperature of oxygen (see Eq.(\ref{Tsp})) for all models. Color coded is the stellar luminosity.}
\label{fig:TgOI_Lstar}
\end{figure}

\begin{figure}
\centering
\includegraphics[width=9cm]{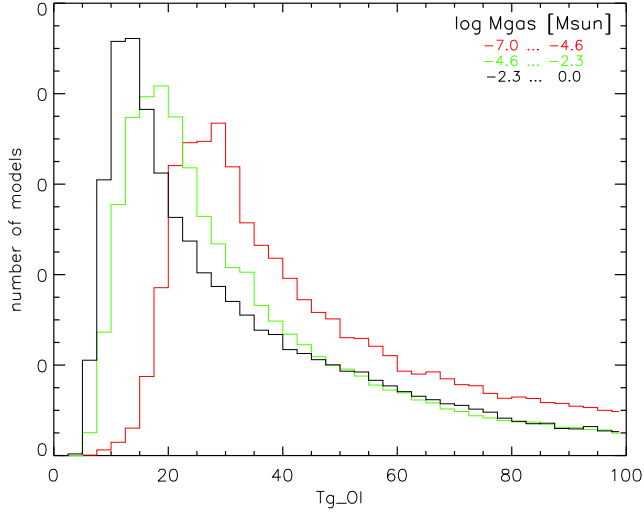}
\caption{Mean gas temperature of oxygen (see Eq.(\ref{Tsp})) for all models. Color coded is the disk gas mass in three bins representing optically thin disk models, disk models in the transition from thin to thick, and optically thick disk models.}
\label{fig:TgOI_Mgas}
\end{figure}

We calculate average gas temperatures for various species from the models following 
\begin{equation}
\langle T_{\rm g}^{\rm sp} \rangle = \frac{m_{\rm sp}}{M_{\rm sp}} \int n_{\rm sp}(r,z) T_g(r,z) dV \,\,\,
\label{Tsp}
\end{equation}
where $n_{\rm sp}$ is the species density and $m_{\rm sp}$ its mass. $M_{\rm sp}$ is the total mass of a species in the disk model. The oxygen gas temperature distribution peaks at 20~K and the median gas temperature is 40~K. The distribution shows that $\sim 60$\% of the models stay between 15 and 70~K. The temperature distributions for specific subsamples of parameters, such as the the stellar luminosity and disk gas mass, are shown in Fig.~\ref{fig:TgOI_Lstar} and \ref{fig:TgOI_Mgas}. If we split the sample by luminosity, the higher luminosity models show on average higher gas temperatures. This can be directly attributed to the higher energy deposition rate which leads to an increased gas heating. Looking at low mass disk models (Fig.~\ref{fig:TgOI_Mgas}), we note that their average gas temperature is almost a factor three higher than that of high mass disk models. This is explained by optical depth effects: in models with moderate optical depth, a larger fraction of the total mass can be heated by irradiation. Hence, models with lower disk mass tend to have on average higher gas temperatures. These general trends are based on a huge number of disk models and the actual distribution of gas temperatures is fairly wide with a long shallow tail towards very high gas temperatures (Fig.~\ref{fig:TgOI_Mgas}). 

The O\,{\sc i} gas temperature statistics (and emerging line fluxes) have already been used in \citet{Pinte2010} to identify the need for an extra heating mechanism from the [O\,{\sc i}] line fluxes of T Tauri stars. In that particular case, the stellar luminosity of the low mass stars does not provide enough energy to explain the observed flux levels. Our models show that a UV excess ($f_{\rm UV} = 0.1$), which can originate either from a chromosphere or from ongoing accretion, can provide this extra energy and reproduce the observed fluxes. X-ray heating is an alternative process for T Tauri disks \citep{Nomura2007,Gorti2008,Aresu2011}.

%\begin{figure}
%\centering
%\includegraphics[width=9cm]{TgO_overview.eps}
%\caption{Mean gas temperature of oxygen (see Eq.(\ref{Tsp}) for all models. Color coded are the surface density power law exponent (upper left panel), $f_{\rm UV}$ (upper right panel), stellar luminosity (lower left panel) and disk gas mass (lower right panel).}
%\label{fig:TgO_overview}
%\end{figure}

\begin{figure}
\centering
\includegraphics[width=9cm]{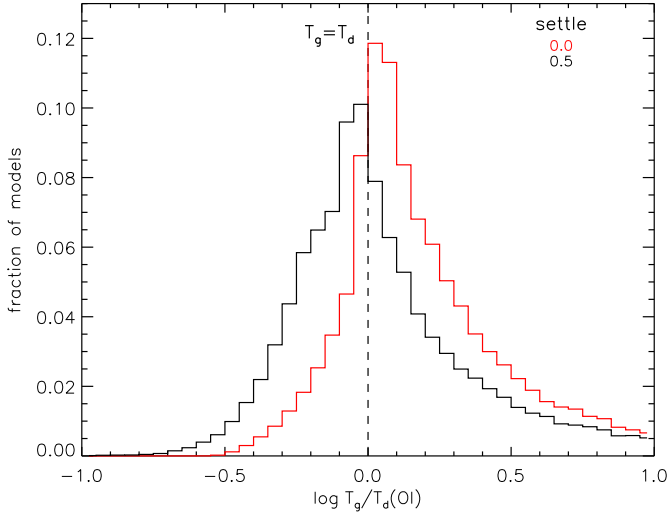}
\caption{Ratio of gas and dust temperatures of neutral oxygen as a function of the dust settling parameter ($s=0$ - without settling, $s=0.5$ - with settling). This ratio is a measure of the temperature de-coupling in the disk surface.}
\label{fig:TgOTdO_settle}
\end{figure}

Fig.~\ref{fig:TgOTdO_settle} shows the impact of dust settling on the coupling of gas and dust temperature for the regions in which atomic oxygen dominates. Since atomic oxygen is widely distributed in the disk, we can take this to first order as an estimate of the amount of thermal gas-dust de-coupling. Well mixed models, $s=0$, show a fairly narrow peak around $T_g=T_d$ and a tendency to have gas temperatures higher than dust temperatures (often in the form of a hot surface gas layer). The settled models, $s=0.5$, show on average gas temperatures that are lower than those of the dust. This is not due to the presence of cool surface layers in settled models (see Sect~\ref{Sect:coolsurfaces}); the mass averaged O\,{\sc i} gas temperature is hardly affected and so the reason lies in the average dust temperature. In settled models, only the small grains stay behind in the disk surface, while the larger ones settle towards the midplane (see Table~\ref{DENT_disk_parameters}). Hence, the effective grain size in the surface of settled models is smaller compared to non-settled models, thereby decreasing the dust emissivity at long wavelength. This leads to slightly warmer dust.

\subsection{CO gas temperatures}

%\begin{figure}
%\centering
%\includegraphics[width=9cm]{TgCO_overview.eps}
%\caption{Mean gas temperature of CO (see Eq.(\ref{Tsp}) for all models. Color coded are the outer radius (upper left panel), $f_{\rm UV}$ (upper right panel), stellar luminosity (lower left panel) and surface density power law exponent (lower right panel).}
%\label{fig:TgCO_overview}
%\end{figure}

\begin{figure}
\centering
\includegraphics[width=9cm]{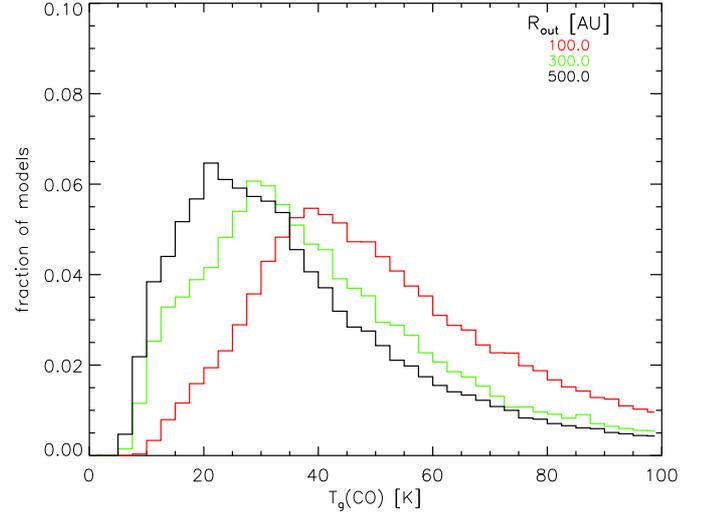}
\caption{Mean gas temperature of CO (see Eq.(\ref{Tsp})) for all models. Color coded is the outer radius $R_{\rm out}$.}
\label{fig:TgCO_Rout}
\end{figure}

\begin{figure}
\centering
\includegraphics[width=9cm]{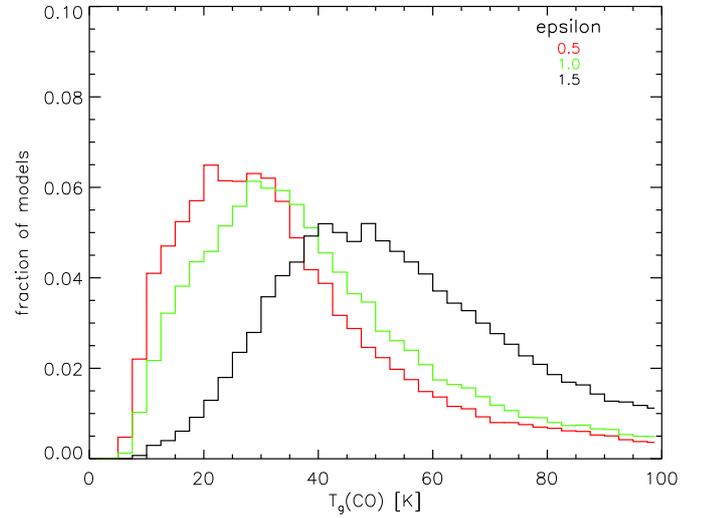}
\caption{Mean gas temperature of CO (see Eq.(\ref{Tsp})) for all models. Color coded is the surface density power law exponent $\epsilon$.}
\label{fig:TgCO_eps}
\end{figure}

Fig.~\ref{fig:TgCO_Rout} and \ref{fig:TgCO_eps} show the distribution of CO mass averaged gas temperatures in the disk models as a function of the outer radius and the surface density power law index. Given our choice of surface density power laws, it is evident that the CO mass is dominated by the material close to the outer radius. Since the gas temperature is also a clear function of distance, models with larger outer radius show on average lower CO gas temperatures. Fig.~\ref{fig:TgCO_Rout} also illustrates the limitations of the concept of a single canonical CO temperature $T_{\rm CO}$ to convert submm line fluxes into disk gas masses. We will get back to this point when looking for possibilities to estimate disk gas masses from line fluxes in Sect.~\ref{estimateMgas_COline} and \ref{estimateMgas_OIandCOline}. The surface density power law exponent also impacts the distribution of mass in the disk. Hence, models with a low $\epsilon$ put more mass in the outer disk compared to the inner disk; hence, they show on average lower CO gas temperatures. In addition, the trends with luminosity and disk mass found for oxygen, hold also for CO.

\begin{figure}
\centering
\includegraphics[width=9cm]{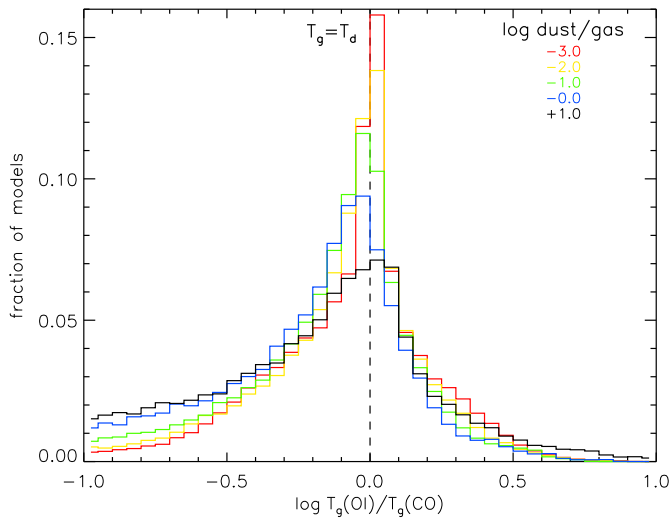}
\caption{Difference between average gas temperatures of atomic oxygen and CO for all models. Color coded is the dust-to-gas mass ratio.}
\label{fig:TgOITgCO_delta}
\end{figure}

For a large fraction of models, the average gas temperatures for atomic oxygen and CO are very similar. Except for the lowest mass disks, both temperatures are within a factor two of each other. This also holds across all gas-to-dust mass ratios as can be seen from Fig.~\ref{fig:TgOITgCO_delta}. 

\subsection{Ionized carbon in disks}

Ionized carbon forms a thin skin on the disk surface with a typical column density of $10^{17-18}$~cm$^{-2}$. Very low mass disk models, $M_{\rm gas} \leq 10^{-5}$~M$_\odot$, are so optically thin that carbon is fully ionized throughout the disk resulting in a constant $M({\rm C^+})/M_{\rm gas}$ fraction of $\sim 1.2 \times 10^{-3}$. This is only weakly dependent on detailed disk parameters.

\begin{figure}
\centering
\includegraphics[width=9cm]{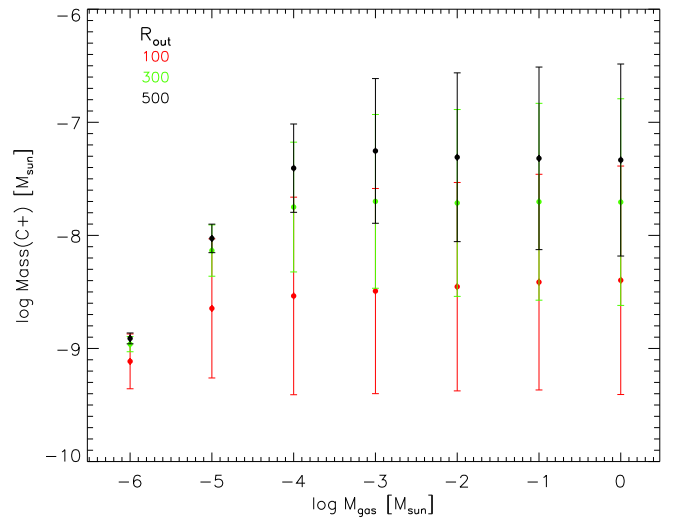}
\caption{Mass of ionized carbon as a function of total disk gas mass. Color coded is the outer disk radius.}
\label{fig:MCII_Mgas_Rout}
\end{figure}

As soon as the ionizing UV radiation is blocked by sufficient dust opacity, carbon turns atomic and subsequently molecular. The higher mass models show thus a decreasing fractional C$^+$ mass, with the total $M({\rm C^+})$ leveling off at $\sim 10^{-8}$~M$_\odot$. The variation between models can be an order of magnitude mostly depending on the choice of disk extension and stellar luminosity (Fig.~\ref{fig:MCII_Mgas_Rout}):  the C$^+$ mass is higher for larger outer disk radii and higher luminosity stars. The gas residing in the outer disk is generally more optically thin --- also to interstellar UV radiation --- than the gas in the inner disk regions.

\begin{figure}
\centering
\includegraphics[width=9cm]{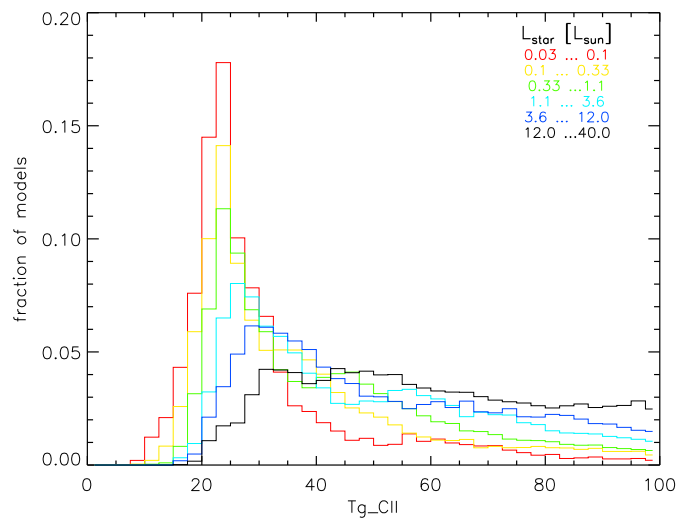}
\caption{Mean gas temperature of ionized carbon (see Eq.(\ref{Tsp})) for all models. Color coded is the stellar luminosity.}
\label{fig:TgCII_Lstar}
\end{figure}

%\begin{figure}
%\centering
%\includegraphics[width=9cm]{TgCII_overview.eps}
%\caption{Mean gas temperature of C\,{\sc ii} (see Eq.(\ref{Tsp}) for all models. Color coded are the outer radius (left panel), and the stellar luminosity (right panel).}
%\label{fig:TgCII_overview}
%\end{figure}

The dependency of the C$^+$ gas temperatures on stellar luminosity is less pronounced than for atomic oxygen. The distribution has a strong peak at low temperatures around $20$~K except for luminosities $L_\ast > 3.6$~L$_\odot$ (Fig.~\ref{fig:TgCII_Lstar}). There is a clear dependency on the outer radius similar to CO, with large models ($R_{\rm out}=300, 500$~AU) having an average C$^+$ gas temperature of $\langle T_{\rm g}^{\rm C^+} \rangle \sim 25$~K. The  small models ($R_{\rm out}=100$~AU) show a much broader flat distribution between 25 and 100~K. The reason is that the C$^+$ mass averaged temperature is dominated by the outermost regions --- just as for CO --- since it is evenly distributed over the entire disk surface. Hence, the inner radius plays no significant role here.

\subsection{Cool surface layers}
\label{Sect:coolsurfaces}

A subset of {\sf DENT} grid models shows very cool surface layers on top of warmer midplanes, i.e.\ models with a positive gas temperature gradient (as seen from the surface). These are very flat, massive disks in which the grains have started to settle. Typical models in this category show $M_{\rm gas} \geq 0.01$~M$_\odot$, no UV excess, settled dust ($s=0.5$) and are non-flaring ($\beta = 0.8, 1.0$). These models appear here for the first time because of the wide unbiased parameter study. 
%Given that we do not iterate on the vertical structure of the disks, how realistic are these configurations?

A related effect has been seen earlier in models by \citet{Aikawa2006} who studied grain growth in disk models that do solve iteratively for the vertical hydrostatic equilibrium. For well mixed models with a large maximum grain size $a_{\rm max}$, those authors find a similar effect. Photoelectric heating is diminished in those models resulting in lower gas temperatures in the disk atmosphere. On the other hand, optical depth is reduced as well allowing stellar UV photons to penetrate deeper and heat the gas and dust at intermediate heights. The authors also find that disks with larger $a_{\rm max}$ flare less than those with small molecular cloud like grain size distributions. This is an immediate result of the lower gas temperatures in the surface layers. For their largest maximum grain size $a_{\rm max}=10$~cm, they find that the gas in the disk atmosphere becomes cooler than the dust. In a subsequent paper \citet{Nomura2007} included the effect of grain settling and X-ray heating and show that --- compared to the standard well-mixed model with molecular cloud like grains --- this indeed lowers the gas temperature in the surface and increases the gas and dust temperatures close to the midplane. The gas temperature gradient becomes much flatter, but there is no pronounced positive temperature gradient. Unfortunately we cannot disentangle from their published model the effect of additional gas heating by X-rays and decrease in gas photoelectric heating due to dust growth and settling. 

\citet{Jonkheid2007} have studied a series of four Herbig disks with an increasing gas-to-dust mass ratio due to grain growth and settling. The basic dust disk models are taken from \citet{Dullemond2004}, who use the dust midplane temperature to set the scale height of the disk. \citet{Jonkheid2007} do not iterate the vertical structure based on the gas temperature and so their approach is somewhat similar to our choice of parametrized disk models. They find that a large fraction of the disk gas mass is indeed cooler than the dust and that the highest gas-to-dust mass ratios lead indeed to a positive gradient in the vertical gas temperature.

 In nature, we do not expect such flat, massive disks, with strong settling to be very common. However, the presence of such cool surface layers would have implications for the disk surface chemical composition and line emission. At cold temperatures, neutral-neutral chemistry becomes inefficient and the formation of warm water in the disk surface could be affected. Self-absorption in the dominant cooling line [O\,{\sc i}]\,63~$\mu$m (see Sect.~\ref{OIlineratio}) and temperature gradients as measured from various $^{12}$CO and $^{13}$CO lines could reveal such cool surfaces.

\begin{figure}
\centering
\includegraphics[width=9cm]{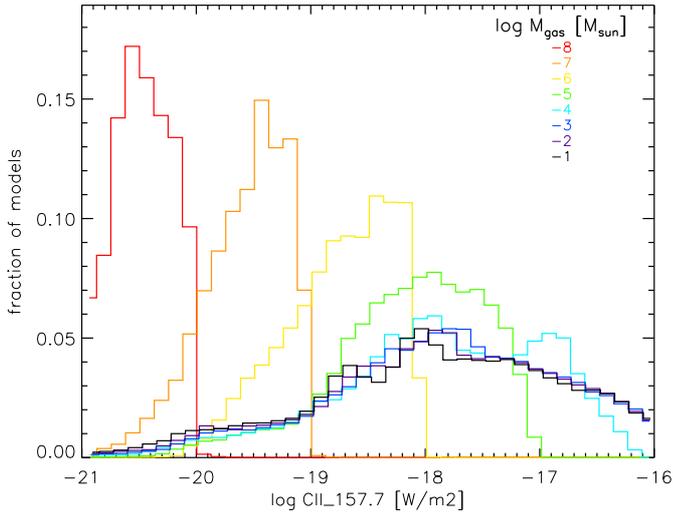}
\caption{Statistical distribution of [C\,{\sc ii}]\,158~$\mu$m emission as a function of disk gas mass. Color codes is the logarithmic disk gas mass.}
\label{fig:CII157}
\end{figure}

\section{Line diagnostics}
\label{diagnostics}

In the following sections, we explore the diagnostic power of individual emission lines from the {\sf DENT} grid. We check especially for clean dependencies on disk input parameters such as outer radius, disk mass, dust-to-gas mass ratio, surface density power law exponent etc.

\subsection{The C$^+$ fine structure line at $158~\mu$m}

%\begin{figure}
%\centering
%\includegraphics[width=9cm]{betaleq1delta0.01-1inclleq60_CII157_Mgas.ps}
%\caption{Statistical distribution of [C\,{\sc ii}]~157~$\mu$m emission as a function of disk gas mass. Selected are models with $\beta \geq 1.0$, $0.01 \leq \delta \leq 1.0$ and inclination $\leq 60^{\rm o}$.}
%\label{fig:CII157}
%\end{figure}

Since ionized carbon originates from a thin skin around the disks, its mass (column densities) tends to be constant until the disk gas mass itself becomes so low that the disk is transparent to UV photons. The total [C\,{\sc ii}]\,158~$\mu$m flux correlates well with the mass in C$^+$ indicating that the line is mostly optically thin. 

\begin{figure}
\centering
\includegraphics[width=9cm]{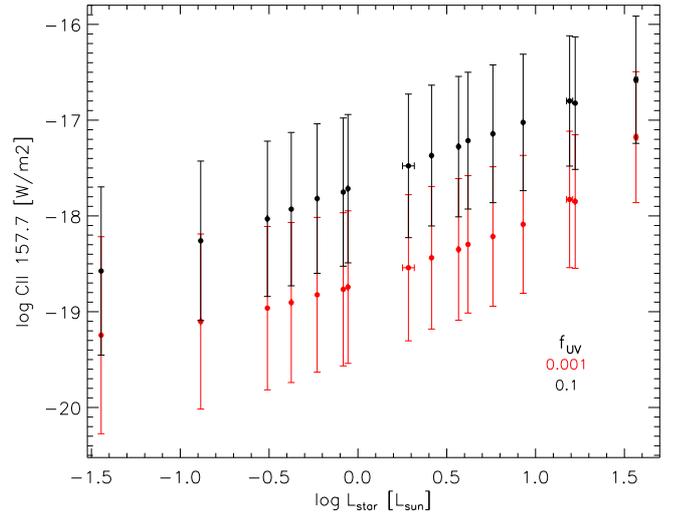}
\caption{[C\,{\sc ii}]\,158~$\mu$m fluxes as a function of stellar luminosity for disk models with $M_{\rm gas} > 10^{-5}$~M$_\odot$. Color coded is the UV excess $f_{\rm UV}$.}
\label{fig:CII157-fUV}
\end{figure}

Fig.~\ref{fig:CII157} illustrates this effect nicely through the statistical distribution of [C\,{\sc ii}]\,158~$\mu$m fluxes as a function of disk mass. Down to disk masses of $10^{-4}$~M$_\odot$, the disks stay optically thick and the C$^+$ mass fraction and hence its emission stays constant around $\sim 3\,10^{-18}$~W/m$^{2}$. The large width of the flux distribution for these models reflects the spread in stellar luminosity, UV excess, outer disk radius, flaring angle; the higher fluxes results from models with high luminosities (stellar and/or UV --- see Fig.~\ref{fig:CII157-fUV}), large outer radii, and flaring geometry. The  [C\,{\sc ii}] line flux depends only weakly on the disk surface density gradient, the gas-to-dust mass ratio, settling, the minimum grain radius $a_{\rm min}$ and the inner disk radius. For lower gas masses, the peak of the [C\,{\sc ii}] distribution shifts towards lower fluxes and the distribution becomes much narrower.

The {\sf DENT} grid was generated using only two values of UV excess, 10\% and 0.1\% of the total stellar luminosity. While these values  bracket the UV excess observed for young T~Tauri stars, it can lead to an overestimate of the UV radiation for Herbig Ae/Be stars that do not possess an excess in the first place. Fig.~\ref{fig:CII157-fUV} shows that the [C\,{\sc ii}] emission increases by an order of magnitude for the high UV models. The Herbig Ae/Be models with pure photospheric UV fluxes would fall even below the red points in Fig.~\ref{fig:CII157-fUV}.

 In practice, the [C\,{\sc ii}]\,158~$\mu$m line is easily contaminated by the low density surrounding gas, even with Herschel/{\sf PACS}. Although the line is potentially a good tracer as shown above, it requires a very careful observational procedure and data reduction to ensure that the flux measured is indeed purely from the disk.

\subsection{Oxygen fine structure lines at $63$ and $145~\mu$m}

In the {\sf DENT} grid, 80\% of the disk models show a $145~\mu$m line that is a factor $10-100$ weaker than the $63~\mu$m line. Given that the Herschel sensitivity limits for both lines are very similar, this clearly favors the detection of the $63~\mu$m line.

\citet{Woitke2010} have shown that the $63~\mu$m fine structure line could be promising as an order of magnitude disk gas mass estimator. Their Fig.~4 illustrates that if we can estimate the average excitation temperature of neutral oxygen, the measurement of the $63~\mu$m line can be used to derive gas masses in the context of a statistical analysis of a large sample of disks such as the {\sc GASPS} sample. 

%\begin{figure}
%\centering
%\includegraphics[width=9cm]{betaleq1delta0.01-1inclleq60_O145_Mgas_TgOI.ps}
%\caption{Disk gas mass versus [O\,{\sc i}]~145~$\mu$m line emission. Selected are models with $\beta \geq 1.0$, $0.01 \leq \delta \leq 1.0$ and inclination $\leq 60^{\rm o}$.}
%\label{fig:OI145}
%\end{figure}

\subsection{The [O\,{\sc i}]\,63/145 line ratio}
\label{OIlineratio}

In PDRs, low [O\,{\sc i}] line ratios ($<10$) are expected if both lines are optically thick \citep{Tielens1985} and the gas is cooler than $\sim 200$~K. \citet{Liseau2006} argue that confusion with cool envelope gas might explain ratios of a few in observations of YSO's. \citet{Lorenzetti2002} suggested that high optical depths in the [O\,{\sc i}]\,63 and 145~$\mu$m lines can explain the [O\,{\sc i}]\,63/145 ratios of $2-10$ observed by ISO/LWS in massive disks around early-type HerbigAeBe stars. 

\begin{figure*}
  \centering
  \begin{tabular}{cc}
  \hspace*{-4mm}\includegraphics[width=94mm,height=84mm]{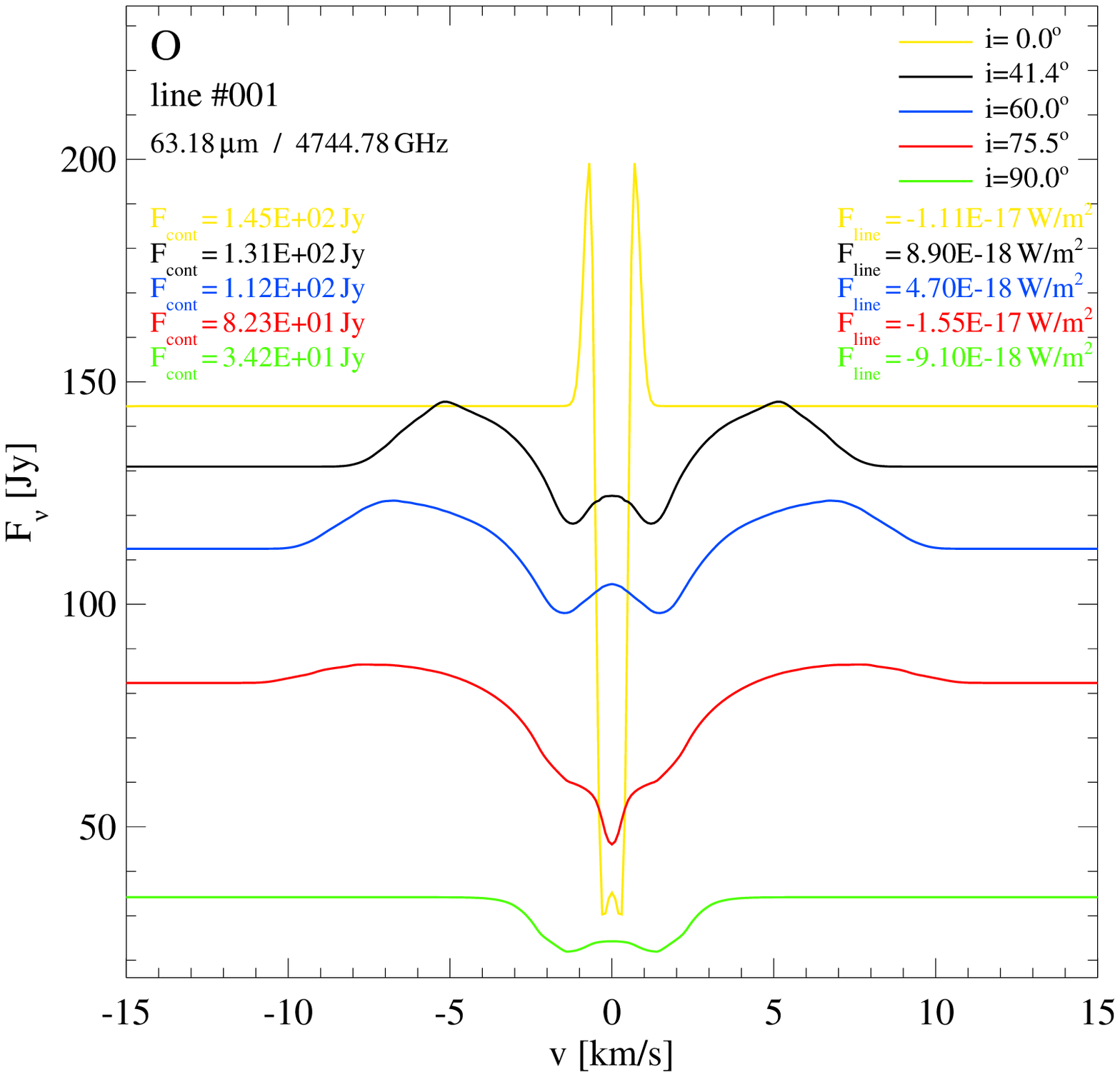}
  \hspace*{0mm}\includegraphics[width=94mm,height=84mm]{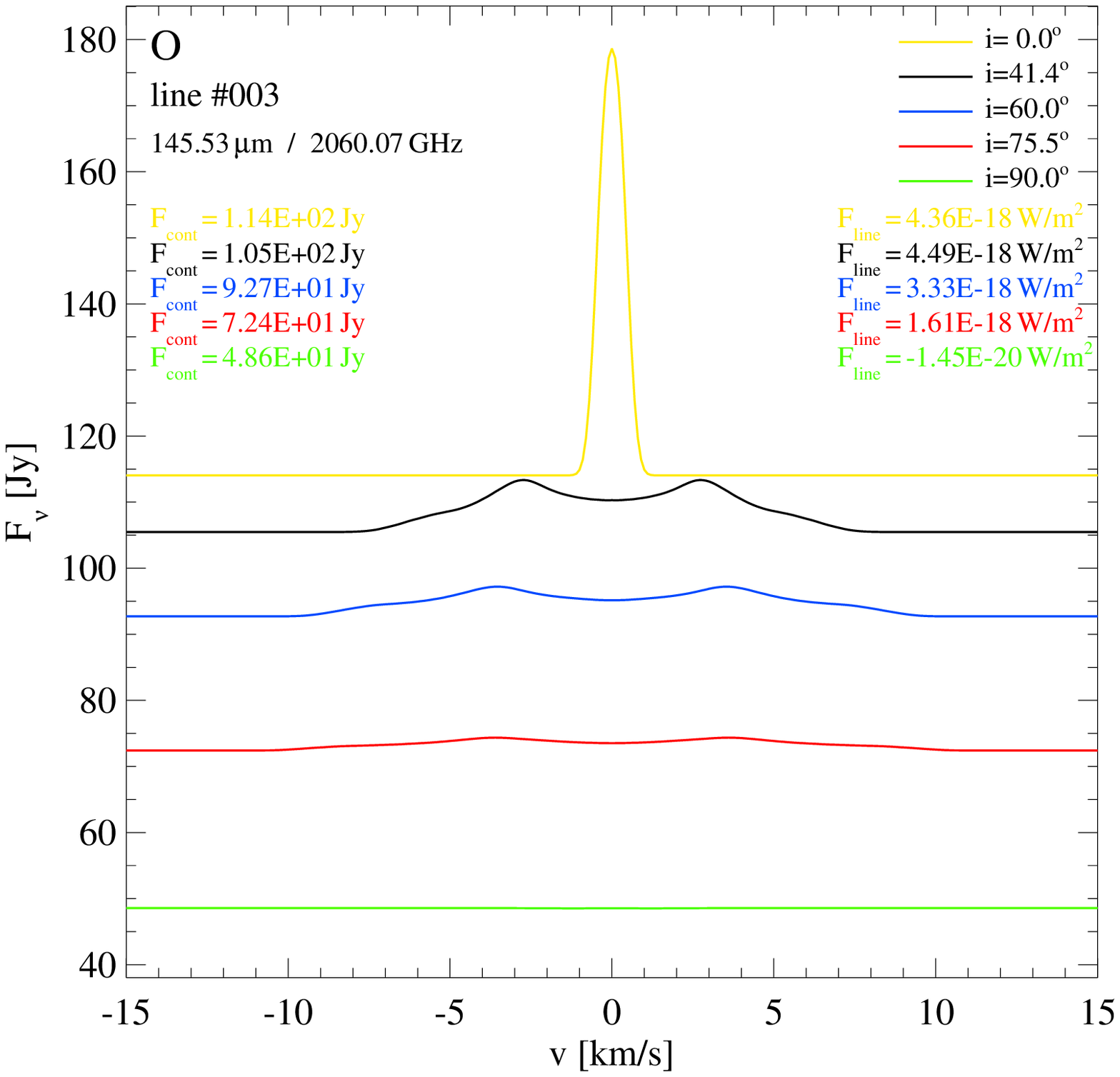}
  \\[-1mm]
  \end{tabular}
  \caption{Unusual behaviour of the [O\,{\sc i}] lines in some disk models. For massive disks with settled dust distribution, the fundamental $\rm[O\,{\sc i}]\,63~\mu$m line may go into central absorption (l.h.s.) whereas the $\rm[O\,{\sc i}]\,145~\mu$m line stays in emission (r.h.s.), which complicates the data interpretation.  Model parameters $M_\star\!=\!2\rm\,M_\odot$, $age\!=\!3\rm\,Myr$, ($T_{\rm eff}\!=\!4960\,$K, $L_\star\!=\!3.66\,L_\odot$), $f_{\rm UV}\!=\!0.001$, $M_{\rm gas}\!=\!0.1\rm\,M_\odot$, $\rm dust/gas\!=\!0.01$, $R_{\rm in}\!=\!17.8\,$AU, $R_{\rm out}\!=\!500\,$AU, $\epsilon\!=\!0.5$, $\beta\!=\!1.0$, $s\!=\!0.5$, $a_{\rm min}\!=\!1\,\mu$m.}
  \label{fig:strangeOI}
  \vspace*{-1mm}
\end{figure*}

The grid results presented here confirm that the [O\,{\sc i}]\,63/145 ratios for massive ($>$10$^{-2}$~M$_\odot$) disks is of the order of $1-10$. In our models, we observe that [O\,{\sc i}]\,$63/145\!<\!10$ can be produced by the disks themselves, when [O\,{\sc i}]\,63~$\mu$m is about to go into absorption, whereas [O\,{\sc i}]\,145~$\mu$m is still in emission. As a fundamental line, [O\,{\sc i}]\,63~$\mu$m can go into central absorption if the emitted light has to pass through another cold neutral oxygen layer above, whereas the [O\,{\sc i}]\,145~$\mu$m line practically never goes into absorption. These effects happen for massive disks only, in particular for settled dust distributions, and are strongly dependent on inclination, see Fig.~\ref{fig:strangeOI} {and also Sect.~\ref{Sect:coolsurfaces}}. The hypothesis that such models exist is a unique testcase for the {\sc SOFIA/GREAT} instrument that can observe the [O\,{\sc i}]\,63~$\mu$m line with high spectral resolution and thus check for self-absorption within the line profile.

\begin{figure}[h]
\centering
\includegraphics[width=9cm]{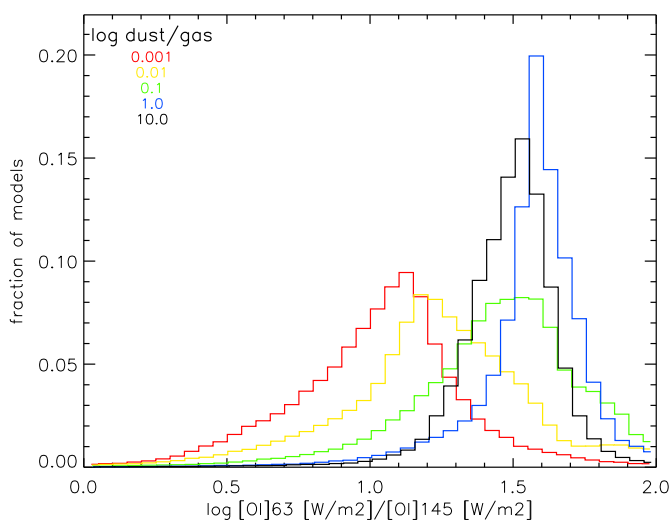}
\caption{Statistical distribution of [O\,{\sc i}]\,63/145 line ratios as a function of the dust-to-gas mass ratio.}
\label{fig:OIlineratio_delta}
\end{figure}

The median [O\,{\sc i}] line ratio for the entire {\sf DENT} grid is 25. Contrary to PDRs, the [O\,{\sc i}]\,63/145 line ratio of disk models is not sensitive to the average oxygen gas temperature over a wide range of temperatures between 50 and 500~K. The line ratio correlates instead with e.g.\ the dust-to-gas mass ratio (Fig.~\ref{fig:OIlineratio_delta}). 
%Thus its diagnostic capabilities are very limited and if at all require a good understanding of the complex interplay of various parameters.

%Insert here text passage from the old letter draft on [O\,{\sc i}]\,63~$\mu$m self-absorption and how it leads to very low line ratios. Include the Lorenzetti et al. and Liseau et al. conclusions and refer to Fig.~\ref{fig:coolsurface}.

%\begin{figure}[h]
%\centering
%\includegraphics[width=9cm]{coolsurface.eps}
%\caption{Tail of the statistical distribution of [O\,{\sc i}]\,63/145 line ratios with various color codes. Selected are only models with inclination $\leq 60^{\rm o}$.}
%\label{fig:coolsurface}
%\end{figure}

\subsection{The CO low J rotational lines}
\label{Sect:COlowJ}

\begin{figure}
%\centering
\includegraphics[width=9cm]{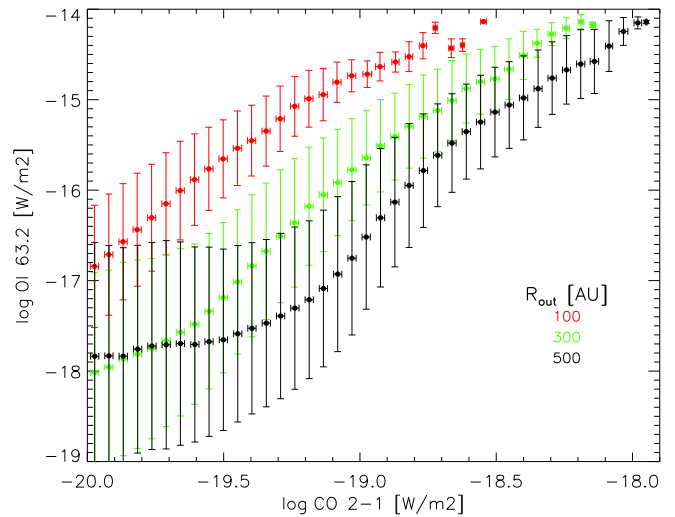}
\caption{[O\,{\sc i}]\,63~$\mu$m versus $^{12}$CO J=2-1 line emission. 
%Selected are models with $\beta \geq 1.0$, $0.01 \leq \delta \leq 1.0$. 
Color coded is the outer disk radius with values of 100, 300 and 500~AU.}
\label{fig:CO2-1}
\end{figure}

The low J $^{12}$CO rotational lines are optically thick and arise mainly from the outer disk surfaces. Hence they form an excellent tracer of the outer disk radius. The larger the disk, the larger the emitting surface. The oxygen fine structure line at 63~$\mu$m is not affected by the disk outer radius for $R_{\rm out} \geq 100$~AU \citep{Kamp2010}. Fig.~\ref{fig:CO2-1} shows the mean [O\,{\sc i}] line flux plotted against the CO~2-1 flux with the outer disk radius coded in color. In this diagram, the disk models clearly separate into small ones with $R_{\rm out} = 100$~AU and ones that are larger than $300$~AU. This indicates that a combination of the oxygen and CO line can help to remove the outer disk radius degeneracy in measuring disk masses.
%The separation is less evident and the scatter becomes larger if using the lower or higher J lines.

\begin{figure}
\includegraphics[width=9cm]{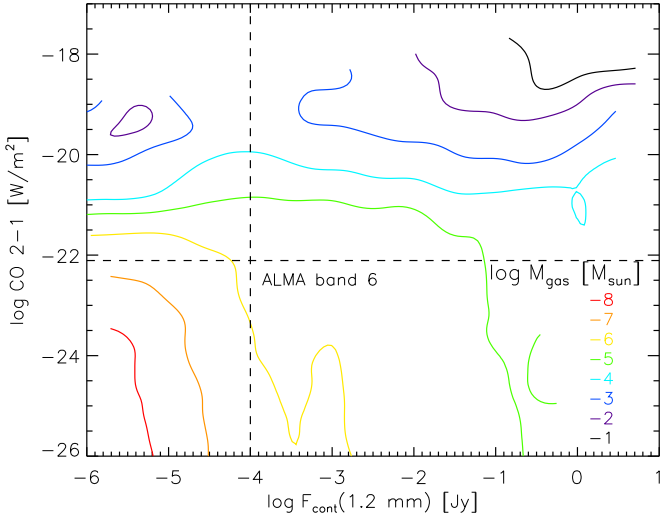}
\caption{$^{12}$CO J=2-1 line emission versus continuum flux at 1.2~mm. For discrete intervals in CO line flux and 1.2~mm flux, mean disk masses are calculated from the ensemble of grid models within that bin. The colored contours show the distribution of these mean disk gas masses. Overplotted are the line and continuum sensitivity limits of ALMA band 6.}
\label{fig:CO2-1-ALMA}
\end{figure}

Detection limits with current radio telescopes such as APEX or SMA are a few times $10^{-20}$~W/m$^2$. With ALMA coming along soon, we will be able to detect much fainter disks down to $\sim 10^{-22}$~W/m$^2$. Fig~\ref{fig:CO2-1-ALMA} shows the CO~2-1 line flux versus the dust continuum at 1.2~mm with the respective ALMA band 6 detection limits. These grid results suggest that ALMA is sensitive enough to detect disk masses of the order of $10^{-6}$~M$_\odot$.

\subsection{The [O\,{\sc i}]\,63/CO~2-1 line ratio}

\begin{figure}
\centering
\includegraphics[width=9cm]{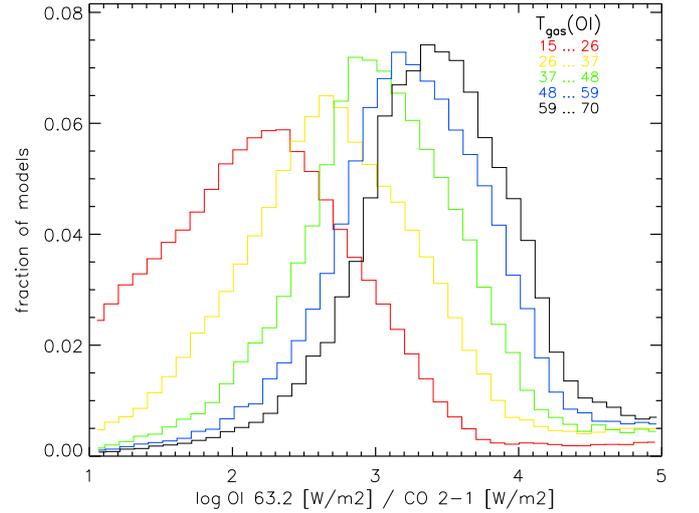}
\caption{Histogram of the [O\,{\sc i}]\,63/CO~2-1 line ratio in models with $15 < \langle T_{\rm g}({\rm O})\rangle < 70$~K (60\% of all models, see Sect.~\ref{Sect:OITg}). Color coded is the mass average oxygen gas temperature $\langle T_{\rm g}({\rm O})\rangle$.}
\label{fig:OI63zuCO21_TgOI_hist}
\end{figure}

\begin{figure}
\includegraphics[width=9cm]{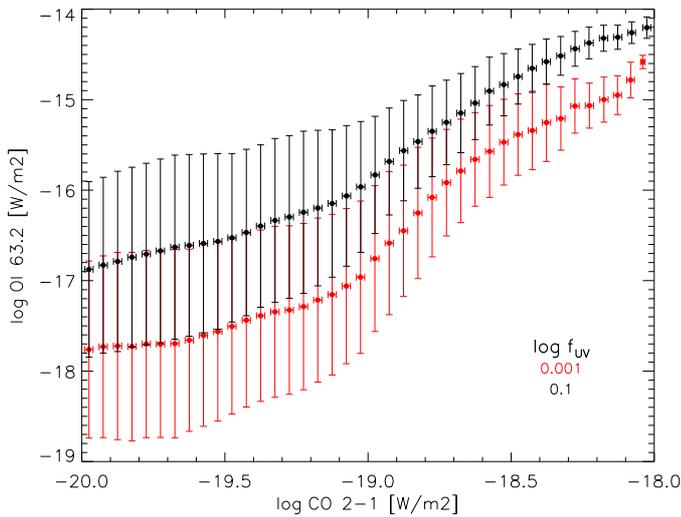}
\caption{[O\,{\sc i}]\,63~$\mu$m versus $^{12}$CO J=2-1 line emission. 
Color coded is the amount of UV excess, $f_{\rm UV}$.}
\label{fig:CO2-1OI63fUV}
\end{figure}

Fig.~\ref{fig:OI63zuCO21_TgOI_hist} illustrates that the [O\,{\sc i}]\,63/CO~2-1 line ratio correlates with the average oxygen gas temperature in the range 15-70~K. If the [O\,{\sc i}]\,63~$\mu$m line is optically thin, it depends on disk gas mass and average oxygen gas temperature \citep{Woitke2010}. The CO~2-1 line is always optically thick (down to $M_{\rm gas} \sim 10^{-4}$~M$_\odot$) and is therefore a function of the CO gas temperature and disk surface area, i.e.\ $R_{\rm out}$. Using the two lines [O\,{\sc i}]\,63~$\mu$m and CO~2-1, we also get a handle of the amount of extra UV radiation irradiating the disk (see Fig.~\ref{fig:CO2-1OI63fUV}).

From Fig.~\ref{fig:CO2-1}, \ref{fig:OI63zuCO21_TgOI_hist} and \ref{fig:CO2-1OI63fUV} it becomes clear that this [O\,{\sc i}]/CO line ratio captures in a complex way a number of disk properties such as the average gas temperature, the disk outer radius, the flaring index and the amount of extra UV irradiation ($f_{\rm UV}$). Thus this line ratio is a powerful means to break the modeling degeneracy and find a suitable method to estimate gas masses even in the absence of a large number of observing constraints. This is especially relevant for large surveys such as {\sf GASPS} (Dent et al.\ 2011 in preparation) that contains many objects where only the [O\,{\sc i}]\,63~$\mu$m fine structure line is detected and ancillary data is mostly restricted to SED's and CO submm lines.

%The very massive disk models show $\langle T({\rm O}) \rangle$ on average a factor three smaller than $\langle T({\rm CO}) \rangle$. This effect is related to the cool surface layers, where we see the  [O\,{\sc i}]\,63~$\mu$m line going in absorption. The parameter combination that promotes this effect best is a non-flaring disk with settled dust and a small inner radius ($R_{\rm in} = R_{\rm subli}$).  NOT TRUE

%Also among the low mass disk models ($M_{\rm g}\leq 10^{-7}$~M$_\odot$), we see a broad tail of models with $\langle T({\rm O}) \rangle < \langle T({\rm CO}) \rangle$. {\bf A parameter combination that favors these temperature ratio is a steep surface density profile ($\epsilon \geq 1.0$) and a large outer radius ($R_{\rm out} \geq 300$~AU).} THIS IS IMPACTED BY LOW MASS MODELS HAVING UNREALISTIC O MASSES - FIX IN IDL SCRIPT DOES NOT ALLOW US TO TRUST THESE MASSES SO ALSO THE <TGOI> IS BIASED TOWARDS MATERIAL THAT HAS BEEN GLUED TO THE DISK BY THE MIN DENSITY CRITERIUM
% those models contain large amounts of CO, because the mass distribution in these disks allows for sufficient self-shielding \remark{check this with CO mass plots}.

\subsection{Water lines}

Due to the molecular structure, water emission lines are not only difficult to calculate (non-LTE, population inversion, optical depth effects), but also difficult to interpret in terms of global disk parameters such as gas mass or surface density profiles \citep{Cernicharo2009}. \citet{Woitke2009b} traced the origin of water lines in the disks around Herbig Ae stars back to three main water reservoirs: (1) a main water reservoir in the central midplane out to $\sim 10$~AU (containing most of the water mass), (2) a cold water belt at $r \approx 10-100$~AU and $z/r \approx 0.05$ (originating from desorption of water ice mantles) and (3) a hot water surface layer at distances of $2-25$~AU (originating from an active neutral-neutral chemistry at $T > 200$~K). Water lines that originate deeper in the disk suffer from the problem that gas and dust temperature couple at some fiducial depth, making it potentially impossible to see the line emission above the strong dust continuum \citep{Woitke2009b,Pontoppidan2010}. 

\begin{figure}
\centering
\includegraphics[width=9cm]{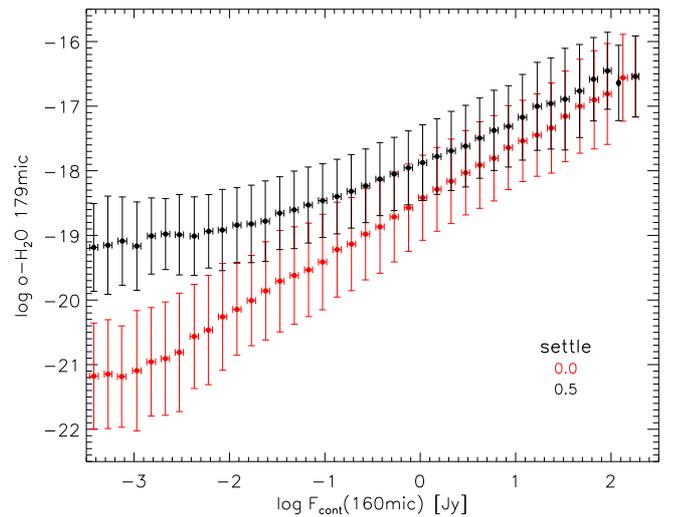}
\caption{Ortho-H$_2$O~$179~\mu$m line flux versus continuum $160~\mu$m flux for massive disk models with $M_{\rm gas} \geq 0.01$~M$_\odot$. Color coded is the settling parameter: $s=0$ denotes homogeneously mixed models, $s=0.5$ denotes models in which the dust grains have settled to a smaller scale height $H(r,a) \propto H(r)\;(\frac{a}{0.05\,\mu{\rm m}})^{-0.25}$. }
\label{fig:H2O}
\end{figure}

%The effect of dust settling is an overall warmer dust temperature averaged over the neutral oxygen in the disk (Eq.\ref{Tsp}). This is due to the lower opacity in the disk surface layers resulting from concentrating the dust mass more towards the midplane. A consequence is also a generally lower optical depth in the vertical direction, thus helping the water lines to shine over the strong dust continuum. Fig.~\ref{fig:H2O} shows that the $210~\mu$m continuum flux is on average half a magnitude smaller in the case with dust settling. However, the distribution of fluxes from individual water lines from settled models is very similar to that from non-settled models (this holds for all ortho- and para-lines that are included in the grid). Hence, the line-to-continuum ratio increases in settled disk models.
Fig.~\ref{fig:H2O} shows that for massive disks ($M_{\rm gas} \geq 0.01$~M$_\odot$), the water lines are generally stronger in the settled models. The effect of dust settling on the water chemistry and line emission is very complex as it affects the available dust surface area, the temperature gradients and the continuum optical depth. Disentangling those effects simply from the grid observables is impossible. Hence, we discuss here two exemplary models to illustrate the general trend displayed in Fig.~\ref{fig:H2O}. Picking a settled and non-settled flaring disk model with $a_{\rm min}=1.0~\mu$m ($M_\ast=2$~M$_\odot$, $T_{\rm eff}=4960$~K, $3.7$~L$_\odot$, $R_{\rm in, out} = 17, 500$~AU, $M_{\rm gas}=0.1$~M$_\odot$, $\delta=0.01$, $\epsilon=0.5$, $\beta=1.0$), we find that the dust in the surface layers of the settled model is warmer on average, while the gas temperature barely changes (using Eq.\ref{Tsp} for oxygen, it changes from 13 to 16~K). This was already discussed in Sect.~\ref{Sect:OITg}. 

%Due to the shallow surface density slope, the average temperature is biased towards the outer cooler regions. Since our grid was calculated without PAHs, the dominant heating mechanism is not photoelectric heating but XXX. 

Even though the temperatures are low, there is very little grain surface area available for e.g.\ H$_2$ or ice formation. The H$_2$ formation on grain surfaces is thus inefficient and atomic hydrogen abundances are high in these settled models. As a consequence of the lack of grain surfaces in the settled model, molecular hydrogen forms via the H$^-$ route and the atomic hydrogen abundance stays a factor 100 higher than in the non-settled models. This leads to an efficient cold water formation route driven by radiative association
\begin{eqnarray}
{\rm H + O} & \rightarrow & {\rm OH + h\nu} \\
{\rm H + OH} & \rightarrow & {\rm H_2O + h\nu}
\end{eqnarray}
The relevance of radiative association for cool atomic H regions was already pointed out by \citet{Glassgold2009}, but found to be irrelevant for the inner disk structures that they studied. These reactions have low rate coefficients\footnote{The UMIST rates are $k({\rm H+O})=9.90\,10^{-19} (T/300~{\rm K})^{-0.38}~{\rm cm}^3~{\rm s}^{-1}$, and $k({\rm H+OH})=5.26\,10^{-18} (T/300~{\rm K})^{-5.22} \exp(90/T)~{\rm cm}^3~{\rm s}^{-1}$ \citep{Field1980}}, but with atomic hydrogen being a factor 100 more abundant in settled models, they dominate the formation route of water. The surface layers ($z/r > 0.2$) of the settled model show water abundances that are several orders of magnitude higher than those of the unsettled model.
%A consequence of settling is a lower vertical optical depth in the disk surface and hence more UV penetration at the same height.}
%, thus helping the water lines to shine over the strong dust continuum. Fig.~\ref{fig:H2O} shows that the $210~\mu$m continuum flux is on average half a magnitude smaller in the case with dust settling. However, the distribution of fluxes from individual water lines from settled models is very similar to that from non-settled models (this holds for all ortho- and para-lines that are included in the grid). Hence, the line-to-continuum ratio increases in settled disk models.

%Settled disk models with $T_{\rm gas}\!>\!T_{\rm dust}$ fit the mid-IR Spitzer observed line-to-continuum ratios of T~Tauri disks much better than well mixed disk models as noted by \citet{Meijerink2009}. These mid-IR lines have larger excitation temperatures and originate generally closer to the star compared to the $179~\mu$m line discussed above ($T_{\rm ex} = 114$~K) and a more quantitative comparison with their results is hence not possible at this stage.

\section{Diagnostic tools}

In the following, we discuss a few methods to measure the disk gas masses from gas line observations. It is evident by the way how we set up the parameter space that the {\sf DENT} grid does not reflect a proper distribution of disk properties, e.g.\ number of T Tauri versus Herbig disks, fractional distribution of disk masses reflecting disk lifetimes etc.. Hence, our aim is not to decide which method is better, but to put the various methods into the context of disk modeling and discuss some of the limitations and uncertainties that become evident if one confronts the simple methods with a wide variety of disk appearances (sizes, flaring indices, gas-to-dust mass ratios, etc.).

The most common technique to estimate disk gas masses has been the use of the mm dust continuum flux. At those wavelengths, the disk is assumed to be optically thin. Hence, the mm-flux should reflect the entire dust mass contained in large grains. The main uncertainties in converting the mm-flux into a total dust and subsequently gas mass are the dust opacities $\kappa_\nu \sim \kappa_0 (\nu/\nu_0)^{-\beta}$ and the dust-to-gas mass ratio $\delta$. An average opacity index $\beta \sim 0.46$ has been recently found for the disks in the $\rho$ Ophiuchus star forming region \citep{Ricci2010}; these low values of $\beta$ (compared to 1.7 for the dust in the interstellar medium) are generally interpreted in terms of grain growth, see also \citet{Draine2006}. However, recent interferometric data suggests that $\beta$ varies with location in the disk \citep[e.g.][]{Isella2010}. The dust-to-gas mass ratio is generally assumed to be the canonical value of $\delta=0.01$ found for the interstellar medium neglecting potential differences in dust and gas evolutionary timescales. Even if the dust mass could be estimated very precisely, this method would still leave us will large uncertainties on $M_{\rm gas}$ because of the unknown conversion factor. Hence, the intrinsic uncertainties of this method are much larger than an order of magnitude. 

Very few direct estimates of the gas-to-dust mass ratio exist for disks. An example is the study by  \citep{Glauser2008} that estimates dust and gas mass on the same disk line-of-sight in a disk. The derived ratio of $220^{+170}_{-150}$ is compatible with the canonical value of 100.

The following three sections discuss potential methods only based on gas emission line fluxes, their limitations and uncertainties.

\subsection{[O\,{\sc i}] and [C\,{\sc ii}] fine structure line ratios}

\begin{figure}
\centering
\includegraphics[width=9cm]{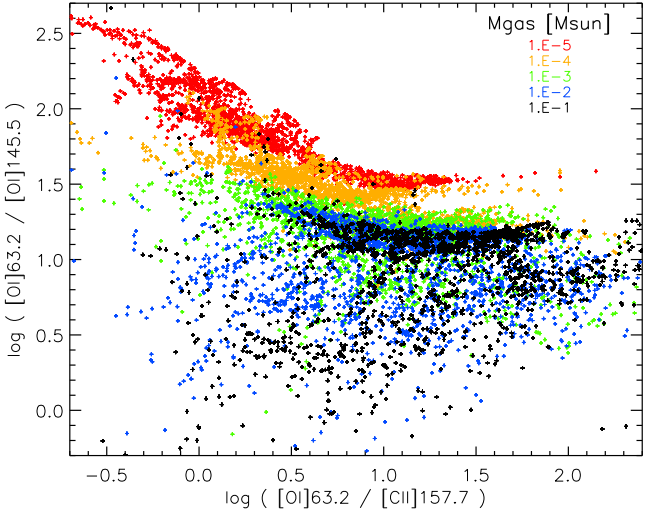}
\caption{Two colour line flux diagram, [O\,{\sc i}]\,63/145 versus [O\,{\sc i}]\,63/[C\,{\sc ii}]\,158, of a sub-selection of 10740 {\sf DENT} models with $\rho_{\rm d}/\rho_{\rm g}\!=0.01$, $\epsilon\!=\!0.5$, $s\!=\!0.5$, inclination angle $41^o$. Color coded is the disk gas mass.}
\label{fig:LineRatios}
\end{figure}

Figure~\ref{fig:LineRatios} shows a two-colour line flux diagram suggesting a method how to determine the disk gas mass purely from three line flux observations.  As was argued by \citet{Kamp2010}, the [O\,{\sc i}]\,63/145 line ratio shows a clear correlation with gas mass.
%, because the $145~\mu$m line has a larger critical density, hence decreases more rapidly with decreasing disk mass than the $63~\mu$m line. 
This plot includes the results for all types of central stars, low and high excess UV, flared and non-flared disks, continuous disks as well as disks with large inner holes, and varying dust parameters and outer disk radii. Despite this variety of disk types, all models with a certain disk gas mass fall into certain regions in this two-colour diagram. Apparently, the main dependencies on $f_{\rm UV}$ and $\beta$ tend to cancel out here, because these parameters shift all line fluxes up and down equally.

However, the diagram looks different for other sub-selections of models as the one stated in the caption of Fig.~\ref{fig:LineRatios}. A steeper surface density power law (higher $\epsilon$) weakens in particular the [C\,{\sc ii}]\,158~$\mu$m flux and hence shifts the points rightward in the figure. 

Selecting only settled models has shifted the location of massive models ($M_{\rm gas} \geq 0.01$~M$_\odot$) downward in Fig.~\ref{fig:LineRatios} compared to a similar plot for unsettled models. The reason for this is mainly [O\,{\sc i}]\,63~$\mu$m self-absorption in the cool surface layers as discussed in Sect.~\ref{Sect:coolsurfaces}.
% In settled models, the dust-to-gas ratio in the {\bf surface} layers is smaller than the canonical $0.01$, and the total dust surface per H-nucleus is smaller; this causes a thermal de-coupling of {\bf gas and dust} in the upper layers. 
%Therefore, dust settling amplifies the shielding effects \remark{How? Explain!}.

In addition, there are other parameters such as the outer disk radius $R_{\rm out}$ and the dust size parameters, which introduce additional non-trivial dependencies and also a large scatter in the various mass bins. Thus a mass determination exclusively based on these three line ratios would implicate large errors especially for the higher mass disk models.

\subsection{Low J CO lines}
%\subsection{The $^{12}$CO~3-2 line}
\label{estimateMgas_COline}

For many objects, the [O\,{\sc i}]\,145 and [C\,{\sc ii}]\,158~$\mu$m lines will not be available as they are often at least a factor 10 fainter than the [O\,{\sc i}]\,63~$\mu$m line. Therefore, we study in the following also the CO low J lines that are accessible from the ground. Especially with ALMA coming up, these lines will be detectable for a large number of disks (see Sect~\ref{Sect:COlowJ}).

The low J $^{12}$CO lines are often optically thick and can hence provide at most a lower limit to the total gas mass. \citet{Dent2005} present a formula to convert the CO~3-2 line flux $I_{\rm CO\,3-2}$ into a gas mass assuming LTE and a constant $^{12}$CO/H$_2$ conversion factor of $5 \times 10^{-5}$. We adopt here their approach and derive a relation between gas mass and intensity
\begin{equation}
M_{\rm gas} \approx 10^{-4} \left( \frac{T_{\rm ex} + 0.93}{e^{-16.74/T_{\rm ex}}} \right) \left( \frac{d}{\rm 100~pc}\right)^2 I_{\rm CO\,3-2} 
\label{Eq:Dent}
\end{equation}
where $I_{\rm CO\,3-2}$ is the integrated CO~3-2 line intensity in K~km/s. Since most of the CO mass will reside in the outer disk, the line flux will be dominated by cold gas. Hence, an average gas excitation temperature of $T_{\rm ex} = 50$~K is assumed in the following estimates. To apply Eq.(\ref{Eq:Dent}) to our grid data, we need to convert the modeled integrated line fluxes in W/m$^{2}$ to K~km/s \citep[formula adapted from][]{Thi2004}
\begin{equation}
I_{\rm CO\,3-2} = 0.01 \times F_{\rm CO\,3-2} \frac{\lambda_{\rm cm}^3}{2 k} \Omega^{-1} \,\,\, .
\label{Eq:Thi}
\end{equation}
Here, $\lambda = 8.6696 \, 10^{-2}$~cm and $k$ is the Boltzman constant in erg/K. The solid angle is assumed to be $\Omega = \pi \left( {\rm HPBW}/2 \right)^2$, with ${\rm HPBW}=13.7"$ the half power beamwidth of the JCMT at 345~GHz. The formulas for the CO~2-1 line look similar; however, the line is generally a factor 2.6 weaker than the CO~3-2 line reflecting the difference in the Planck function $(\nu_{3-2}/\nu_{2-1})^3$.

\begin{figure}
\includegraphics[width=9cm]{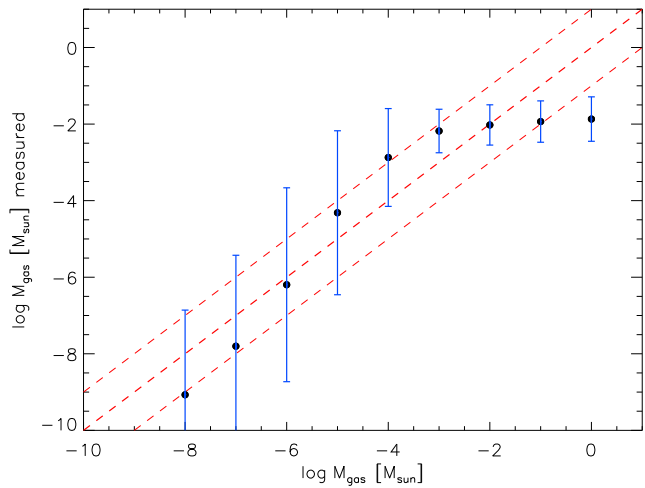}
\caption{Correlation between the gas mass estimated from the modeled CO 3-2 line fluxes (Eq.(\ref{Eq:Dent}) and (\ref{Eq:Thi})) and the original input gas mass of the models. The blue dots are the mean estimated gas masses and the error bar denotes the standard deviation. The red solid line is the expected correlation together with the dashed lines that denote an uncertainty of one order of magnitude.}
\label{fig:Mgas_MgasCOmeasured}
\end{figure}

We insert our modeled fluxes first into Eq.(\ref{Eq:Thi}) and then the converted intensities into Eq.(\ref{Eq:Dent}) to derive a gas mass estimate. These derived values underestimate the actual gas mass, especially at the high mass end, where the CO~3-2 line is optically thick. Fig.~\ref{fig:Mgas_MgasCOmeasured} shows that the behavior is much more complex and that there is a regime, where the simple formula even overestimates the disk gas mass, most likely due to the simplified assumption of a uniform excitation temperature of 50~K for all models. Most notably, the standard deviation across all disk parameters for this method to measure disk gas mass is large, generally more than two orders of magnitude. 

\subsection{The [O\,{\sc i}]\,63/CO~2-1 line ratio}
\label{estimateMgas_OIandCOline}

\begin{figure}
\includegraphics[width=9cm]{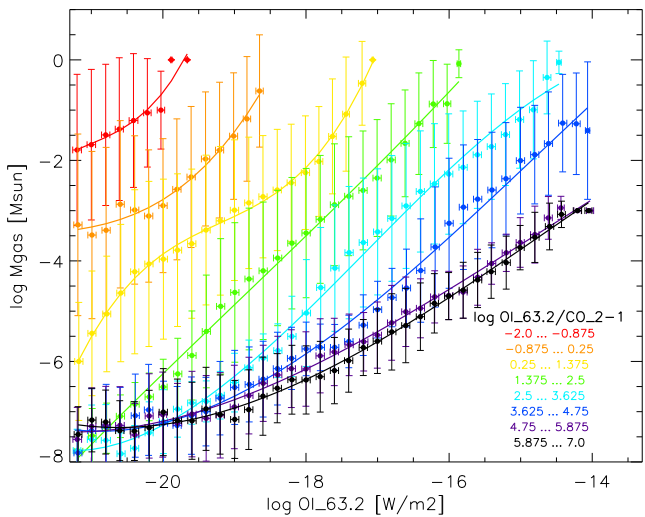}
\caption{Correlation between the gas mass in the disk and the [O\,{\sc i}]\,63~$\mu$m line. The second relevant parameter, the average gas temperature, is color coded through the [O\,{\sc i}]\,63/CO~2-1 line ratio. A selection is made for CO~2-1 fluxes larger than $10^{-23}$~W~m$^{-2}$. The colored lines are polynomial fits of 3rd order to the data.}
\label{fig:Mgas_OI63_OI63zuCO21_means}
\end{figure}

We proceed thus to suggest here an alternative approach using the insight on the diagnostic power of the [O\,{\sc i}]\,63/CO~2-1 line ratio from the previous sections.
%on $\langle T({\rm O}) \rangle$ from the entire {\sf DENT} grid. The basic idea was originally presented in \citet{Woitke2010} and is based on a correlation between the [O\,{\sc i}]\,63~$\mu$m line flux and the disk gas mass. The basic uncertainty there was the average O\,{\sc i} gas temperature. Fig.~\ref{fig:Mgas_OI63_OI63zuCO21_means} illustrates nicely that the [O\,{\sc i}]\,63/CO 2-1 line ratio can now be used to estimate the gas temperature and thus obtain at least an order of magnitude disk mass estimate from the [O\,{\sc i}]\,$63\,\mu$m line flux. 
Fig.~\ref{fig:Mgas_OI63_OI63zuCO21_means} shows for each bin of [O\,{\sc i}]\,63/CO~2-1 line ratios the resulting fit using a 3rd order polynomial. The coefficients for these polynomial fits are presented in Table~\ref{tab:poly}. In the following, we check how accurate the disk masses derived from a combination of these two lines are.

\begin{table}[h]
\caption{Coefficients for the polynomial fit of $M_{\rm gas}$ as a function of [O\,{\sc i}]\,63~$\mu$m flux $F_{\rm [O\,I]}$ and the  [O\,{\sc i}]\,63/CO~2-1 line ratio: $M_{\rm gas} = a_0 + a_1 F_{\rm [O\,I]}  + a_2 F^2_{\rm [O\,I]} + a_3 F^3_{\rm [O\,I]}$. The fits are valid in the range $-21 \leq F_{\rm [O\,I]} \leq -14$~W/m$^2$.}
\begin{center}
\begin{tabular}{c | rrrr}
\hline
$\log$ [O\,{\sc i}]\,63/CO~2-1 &  $a_0$ & $a_1$ & $a_2$ & $a_3$ \\
\hline\\[-2mm]
$-2.0 \ldots -0.875$ & 3446.43 & 490.14 & 23.25 & 0.37 \\
$-0.875 \ldots 0.25$ & 735.29 & 100.93 & 4.60 & 0.07 \\
$0.25 \ldots 1.375$ & 1332.12 & 207.36 & 10.77 & 0.19 \\
$1.375 \ldots 2.5$ &  56.85 & 7.03 & 0.30 & 0.0053 \\
$2.5 \ldots 3.625$ & -123.69 & -23.84 & -1.47 & -0.029 \\
$3.625 \ldots 4.75$ & -21.34 & -6.63 & -0.54 & -0.012 \\
$4.75 \ldots 5.875$ & -13.77 & -3.87 & -0.32 & -0.0072 \\
$5.875 \ldots 7.0$ & -16.98 & -4.76 & -0.39 & -0.0090 \\
\hline
\end{tabular}
\end{center}
\label{tab:poly}
\end{table}%

\begin{figure}
\includegraphics[width=9cm]{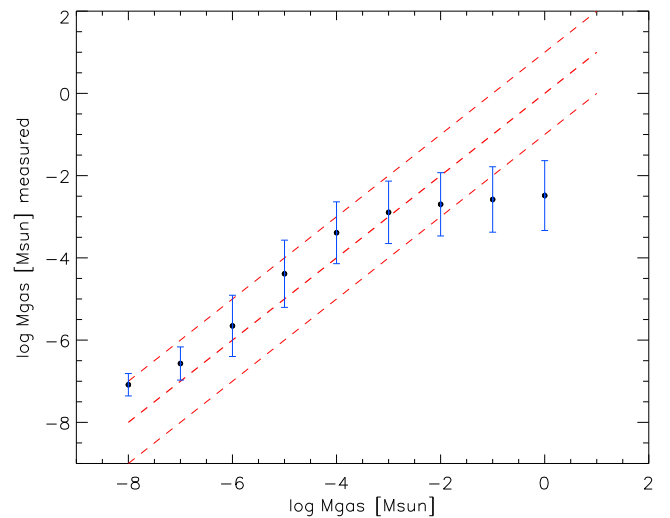}
\caption{Correlation between the gas mass estimated (or 'measured') from a combination of the [O\,{\sc i}]\,63~$\mu$m line flux and the [O\,{\sc i}]\,63/CO~2-1 line ratio and the input gas mass of the disk. The blue dots are the mean `measured' gas masses and the error bar denotes the standard deviation. The red solid line is the expected correlation together with the dashed lines that denote an uncertainty of one order of magnitude.}
\label{fig:Mgasmeasured}
\end{figure}

For every model, we calculate the [O\,{\sc i}]\,63/CO~2-1 line ratio and then select the corresponding curve from Fig.~\ref{fig:Mgas_OI63_OI63zuCO21_means}. This curve in combination with the calculated [O\,{\sc i}]\,63~$\mu$m flux results in a `measured' disk gas mass. This value is then compared to the actual gas mass in the models (Fig.~\ref{fig:Mgasmeasured}). The standard deviation indicates that this method can provide an order of magnitude estimate of the disk gas mass. The method does frequently overestimate the disk gas mass by a factor two. Fig.~\ref{fig:Mgasmeasured} also shows that the relation levels off at disk masses beyond $10^{-3}$~M$_\odot$ where the [O\,{\sc i}]\,63~$\mu$m line becomes optically thick.

For a large statistical sample in which individual disk parameters are less well known, this method can indeed help to assess gas evolution in disks and to measure the gas dispersal timescale. The errors of this method are much smaller than the ones from the CO~3-2 line alone. Of course, more observational constraints on any object should naturally lead to more accurate measurements.

\section{Conclusions}
\label{conclusions}

We have calculated a grid of 300\,000 disk models using the dust Monte Carlo radiative transfer code \mcfost\ and the gas chemistry and heating/cooling balance code \ProDiMo. These models span an 11-dimensional parameter space and enable us to study disk properties such as temperatures, species masses, gas line fluxes and dust continuum fluxes in a statistical manner. The goal is to find suitable gas diagnostics that allow us to characterize the main disk parameters such as radial extent, gas mass and flaring from observational quantities.

The main results found from an extensive statistical analysis of the entire {\sf DENT} model grid are:
\begin{itemize}
%\item average O mass and gas temperature
%\item average CO mass and gas temperature
%\item average C$^+$ mass and gas temperature
\item The [O\,{\sc i}]\,63~$\mu$m traces disk mass but for massive non-flaring, settled disk models without UV excess, the line is optically thick and often suffers from self-absorption in upper cool disk layers.
%\item The [O\,{\sc i}]\,145~$\mu$m traces ...
\item The [C\,{\sc ii}]\,158~$\mu$m fine structure line flux is very sensitive to the stellar UV flux and presence of a  UV excess.
\item The CO low rotational lines trace the outer disk radius.
\item Dust settling enhances the water abundance in the surface layers through an efficient radiative association channel. The initial step of H$_2$ formation proceeds in the gas phase via H$^-$.
\item Low [O\,{\sc i}]\,63/145 line ratios ($<$ a few) can be explained with cool gas in the uppermost surface layers instead of $63~\mu$m absorption by cool envelope remnant gas as suggested by \citet{Liseau2006}.
\item The [O\,{\sc i}]\,63/CO~2-1 line ratio correlates with several disk properties such as the average O\,{\sc i} gas temperature in disks, the outer disk radius, and the UV excess. As such it is a powerful diagnostic to break disk modeling degeneracies.
\item A combination of the [O\,{\sc i}]\,63~$\mu$m flux and the [O\,{\sc i}]\,63/CO~2-1 line ratio can be used for $M_{\rm gas} \leq 10^{-3}$~M$_\odot$ to obtain an order of magnitude estimate for the disk gas mass purely from gas observations. The previously used conversion of a CO submm line flux alone generally leads to larger uncertainties.
\end{itemize}

\begin{acknowledgements}
FM, CP, WFT, and JCA acknowledge PNPS, CNES and ANR (contract ANR-07-BLAN-0221) for financial support.
CP acknowledges funding from the European Comission seventh Framework Program (contracts PIEF-GA-2008-220891 and PERG06-GA-2009-256513). IK acknowledges funding from an NWO MEERVOUD grant.
\end{acknowledgements}

\appendix

\section{The {\sf DENT} grid}

The general outline of our numerical pipeline is described in \citet{Woitke2010} (see their Fig.1). In the first step, \mcfost\ is used to compute the dust temperature $T_{\rm dust}(r,z)$ and frequency dependent radiation field $J_\nu(r,z)$ from radiative transfer. This information is passed on in the second step to \ProDiMo\ to compute the detailed gas chemistry $n_{\rm mol}(r,z)$ and temperature $T_{\rm gas}(r,z)$. The resulting gas structure together with the non-LTE level population numbers of the relevant species is inserted back into \mcfost\ to do the line radiative transfer and calculate the line profiles and integrated fluxes (step 3). In a second radiative transfer step, dust spectral energy distributions (SED) are calculated for all models (step 4).

\begin{table}
\caption{Parameters for the model grid and values assumed.}
\vspace*{-2mm}
\begin{tabular}{lll}
\hline  
\multicolumn{3}{l}{\sf stellar parameter}\\
\hline
{$M_\star$}   
   & stellar mass $[M_\odot]$ & 0.5, 1.0, 1.5, 2.0, 2.5\\
{$age$}
   & age [Myr] & 1, 3, 10, 100\\
{$L_\ast$}
   & dependent parameter & see Table~\ref{DENT_stellar_parameters} \\
{$T_{\rm eff}$}
   & dependent parameter & see Table~\ref{DENT_stellar_parameters} \\
{$f_{\rm UV}$}
   & excess UV $f_{\rm UV}\!=\!L_{UV}/L_\star$ & 0.001, 0.1\\
\hline 
\multicolumn{3}{l}{\sf disk parameter}\\
\hline
{$M_{\rm d}$} 
   & disk dust mass $[M_\odot]$ & 
      \hspace*{0mm}$10^{-7}$, $10^{-6}$, $10^{-5}$, $10^{-4}$,
      $10^{-3}$\hspace*{-2mm}\\
{$\rho_{\rm d}/\rho_{\rm g}$}\hspace*{-2mm} 
   & dust/gas mass ratio $\delta$ & 0.001, 0.01, 0.1, 1, 10\\
{$R_{\rm in}$} 
   & inner disk radius $[R_{\rm subli}]$ & 1, 10, 100\\
{$R_{\rm out}$} 
   & outer disk radius [AU] & 100, 300, 500\\
{$\epsilon$} 
   & column density & \\
   & $N_{\rm H}(r)\propto r^{\displaystyle-\epsilon}$ & 
             0.5, 1.0, 1.5 \\
{$\beta$} 
   & flaring $H(r)=H_0\big(\frac{r}{r_0}\big)^\beta$\hspace*{-3mm} & 
             0.8, 1.0, 1.2\\
{$r_0$}
   & reference radius [AU] & 100 \\
{$H_0$}
   &  scale height at $r_0$ [AU] & 10 \\
{$\chi_{\rm ISM}$}
   & strength of incident ISM UV & 1 \\
{$\zeta_{\rm CR}$}
   & cosmic ray H$_2$ ionization rate & \\
   & [s$^{-1}$] & $5 \times 10^{-17}$ \\
{$f_{\rm PAH}$}
   & abundance of PAHs & \\
   & relative to ISM & 0 \\
{$\alpha$}
   & viscosity parameter & 0 \\
\hline 
\multicolumn{3}{l}{\sf dust parameter}\\
\hline
{$s$} & settling & \\
          & $H(r,a) \propto H(r)\;(\frac{a}{0.05\,\mu{\rm m}})^{-s/2}$ & 0, 0.5\\
{$a_{\rm min}$} 
    & minimum grain size $[\mu m]$ & 0.05, 1\\
{$a_{\rm max}$} 
    & maximum grain size $[\mu m]$ & 1000\\
{$\rho_{\rm gr}$} 
    & grain material density [g/cm$^3$] & 3.5\\
\hline 
\multicolumn{3}{l}{\sf radiative transfer parameter}\\
\hline
{$i$} & inclination & $0^{\rm o}, 41.41^{\rm o}, 60^{\rm o}, 75.52^{\rm o},$ \\
         &                     & $90^{\rm o}$ (edge-on) \\ 
{$v_{\rm turb}$} & turbulent line width [km/s] & 0.15 \\
\end{tabular}
\label{DENT_disk_parameters}
\end{table}

\subsection{Parametrized disk models}

For the shape and mass distribtion of the gas in the disk, we use the following parametric description
\begin{equation}
  \rho(r,z) = \rho_0\,
              \Big(\frac{r}{r_0}\Big)^{-\textstyle\epsilon}\,
              \Big(\frac{H_0}{H(r)}\Big)\,
              \exp\Big(-\frac{z^2}{2\,H(r)^2}\Big)
\end{equation}
between an inner and outer disk radius, $R_{\rm in}$ and $R_{\rm out}$, respectively, with sharp edges. $\rho(r,z)$ is the local gas mass density. The constant $\rho_0$ is adjusted such that the integrated disk mass $2\pi \iint \rho(x,z)\,dz\,r\,dr$ equals $M_{\rm disk}$. $H(r)$ is the vertical scale height of the disk, assuming to vary with radius as
\begin{equation}
  H(r) = H_0\,\Big(\frac{r}{r_0}\Big)^{\,\beta} \ .
\end{equation}
$H_0$ is the reference scale height at reference radius $r_0$. $\epsilon$ is the column density powerlaw index and $\beta$ the flaring power.

\begin{table}[t]
\caption{Stellar parameters of the model grid interpolated from \citet{Siess2000} for solar metallictities.}
\begin{tabular}{llllll}
\hline
$M_\ast$ [M$_\odot$] & Age [Gyr] &	$T_{\rm eff}$ [K] & $R_\ast$ [R$_\odot$] & $L_\ast$ [L$_\odot$] & $\log g$ \\
\hline
  0.50	&	0.001	&	 3771	&	2.13	&	0.93	&	3.5\\
  0.50	&	0.003	&	 3758	&	1.31	&	0.36	&	3.9\\
  0.50	&	0.010	&	 3764	&	0.85	&	0.15	&	4.3\\
  0.50	&	0.100	&	 3828	&	0.43	&	0.05	&	4.9\\
  \hline
  1.00	&	0.001	&	 4278	&	2.62	&	2.34	&	3.6\\
  1.00	&	0.003	&	 4262	&	1.72	&	1.00	&	4.0\\
  1.00	&	0.010	&	 4276	&	1.18	&	0.48	&	4.3\\
  1.00	&	0.100	&	 5357	&	0.90	&	0.66	&	4.5\\
  \hline
  1.50	&	0.001	&	 4536	&	3.11	&	4.17	&	3.6\\
  1.50	&	0.003	&	 4600	&	2.10	&	2.00	&	4.0\\
  1.50	&	0.010	&	 5403	&	1.84	&	2.89	&	4.1\\
  1.50	&	0.100	&	 6872	&	1.44	&	4.63	&	4.3\\
  \hline
  2.00	&	0.001	&	 4710	&	3.60	&	6.47	&	3.6\\
  2.00	&	0.003	&	 4961	&	2.59	&	4.12	&	3.9\\
  2.00	&	0.010	&	 8903	&	1.69	&	17.13 &	4.3\\
  2.00	&	0.100	&	 8829	&	1.65	&	15.94 &	4.3\\
  \hline
  2.50	&	0.001	&	 4853	&	4.13	&	9.59	&	3.6\\
  2.50	&	0.003	&	 5697	&	4.19	&	18.61 &	3.6\\
  2.50	&	0.010	&	10507	&	1.81	&	38.12 &	4.3\\
  2.50	&	0.100	&	10217	&	1.95	&	39.72 &	4.3\\
\end{tabular}
\label{DENT_stellar_parameters}
\end{table}

The dust grains are assumed have a unique powerlaw size distribution 
\begin{equation}
  dn(a) \propto a^{\,-p} da
  \label{eq:dustsizedist}
\end{equation}
between minimum grain radius $\amin$ and maximum grain radius $\amax$. The free constant in Eq.\,(\ref{eq:dustsizedist}) is adjusted to result in the specified dust/gas mass ratio $\rho_{\rm d}/\rho$ \citep[see Sect.\,4.6 in][]{Woitke2009a}, which is assumed to be constant throughout the model volume (\eg no dust settling). The dust absorption and scattering opacities are calculated by applying Mie theory with optical constants for astronomical silicate from \citet{Draine1984}. The {\sf DENT} grid comprises also models with dust setlling, although that particular parameter is only sampled by the two extremes of non-settled models and models in which grains larger than $0.05~\mu$m have settled to a smaller scale height than the gas, $H(r,a) \approx H(r) \left(a/0.05~\mu{\rm m} \right)^{-0.25}$ (for $s=0.5$).

Table~\ref{DENT_disk_parameters} summarizes the free parameters in the {\sf DENT} grid. For completeness, it also contains the fixed parameters $r_0$, $H_0$, $\chi_{\rm ISM}$, $\zeta_{\rm CR}$, $f_{\rm PAH}$, $\alpha$, $a_{\rm max}$, $\rho_{\rm gr}$ and $v_{\rm turb}$.

Effective temperature and luminosity are related parameters and are consistently chosen from the evolutionary tracks of pre-main sequence stars \citep{Siess2000} for solar metallicities. Table~\ref{DENT_stellar_parameters} lists the parameter combinations chosen for the representative five stellar masses 0.5, 1.0, 1.5, 2.0, and 2.5~M$_\odot$. Disk models are calculated at four stellar evolutionary times, 1, 3, 10 and 100~Myr. The stellar input spectra are interpolated from Phoenix stellar atmosphere models \citep{Brott2005} with solar metallicities. 

At UV wavelengths, however, where stellar activity and mass accretion create an UV excess with respect to the classical stellar atmosphere models, we switch to a powerlaw UV input spectrum with spectral
intensity [erg/cm$^2$/s/\AA/sr] 
\begin{equation}
  I_\lambda^\star \propto \lambda^{\,^{\scriptstyle p_{\rm UV}}} \ .
  \label{eq:UVpowerlaw}
\end{equation}
The flux is scaled to yield the prescribed fractional UV luminosity of $f_{\rm UV}\!=\!L_{\rm UV}/L_\star\!=\!0.1$(high UV) or 0.001 (low UV) with the UV luminosity $L_{\rm UV}$ being integrated from 912\,\AA\ to 2500\,\AA\ as introduced by \citet{Woitke2010}. Very few stars have a well known $f_{\rm UV}$ (given the wavelength range defined above and the gaps in observational data of e.g.\ STIS and FUSE). One of the few stars is ET Cha, which has a value of 0.02 (Woitke et al. 2011, submitted).

%The disk models are characterized through power laws
%\begin{eqnarray}
%N_{\rm H}(r) & = & N_0 \left(\frac{r}{r_0}\right)^{-\epsilon} \\
%H(r) & = & H_0 \left( \frac{r}{r_0} \right)^{\beta}
%\end{eqnarray}
%where $N_{\rm H}$ is the total surface density of the disk, $r$ the distance from the star, and $H$ the scale height. The index 0 denotes that this value is specified at the disk inner radius $r_0$. $\epsilon$, $\beta$ and $H_0$ are free parameters and their values can be found along with the other free and fixed parameters in Table~\ref{DENTparameters}. 

\subsection{The observables}

The quantities that can be compared to observations such as the gas line emission and dust SED are computed in step 3 and 4 of the grid modeling procedure (see introduction to Sect.~\ref{DENT}). All quantities are computed for 5 different inclinations that are uniformly distributed, $0^{\rm o}$, $41.41^{\rm o}$, $60^{\rm o}$, $75.52^{\rm o}$ and $90^{\rm o}$ (edge-on). 

The dust SED is computed at $57$ wavelength points between $0.1$ and $3500$~$\mu$m. The line profiles and integrated fluxes of four species are computed: C\,{\sc ii}, O\,{\sc i}, CO and H$_2$O (ortho and para). Table~\ref{linelist} provides an overview of the respective atomic/molecular line data.

\begin{table}
\caption{Atomic and molecular data for lines calculated in the {\sf DENT} grid. All data is taken from the {\sf LAMBDA} database \citep{Schoier2005} with references to the original data indicated in the last column.}
\begin{tabular}{l|llll}
\hline
species     & $\lambda$   & ident. &  $A_{ij}$  & reference \\
            & [$\mu$m]   & &  [s$^{-1}$] &  \\
\hline
C\,{\sc ii} & 157.74    &   $^2$P$_{\frac{3}{2}}$-$^2$P$_{\frac{1}{2}}$  &   2.300(-6) & F77, L77, W02 \\
\hline
O\,{\sc i}  & 63.18     &  $^3$P$_1$-$^3$P$_2$  & 8.865(-5)     &  B98, C80, J92, L77  \\
            & 145.53    &  $^3$P$_0$-$^3$P$_1$   &    1.772(-5)    &  \\
            \hline
$^{12}$CO   &   2600.76 &  J=1-0 & 7.203(-8) & F01, J05, W06, Y10\\
                       &  1300.40  &  J=2-1 & 6.91(-7) & \\
                        &   866.96  &  J=3-2 & 2.497(-6) & \\
                        &  650.25  &   J=4-3 & 6.12(-6) & \\
                        &  520.23  &  J=5-4 & 1.221(-5) & \\
                       & 433.56 &  J=6-5 & 2.137(-5) & \\
                       &  371.65 &  J=7-6 & 3.422(-5) & \\ 
                      & 325.23  &  J=8-7  & 5.134(-5) & \\
                      &  289.12 &  J=9-8 & 7.33(-5) & \\
                      &  260.24 &  J=10-9 & 1.006(-4) & \\
                      & 144.78  &  J=18-17 & 5.695(-4) & \\
                      &  90.16  &  J=29-28 & 2.126(-3) & \\
                      & 79.36 &  J=33-32 & 2.952(-3) & \\
                      &  72.84 &  J=36-35 & 3.638(-3) & \\ 
\hline
o-H$_2$O    &  538.29  &  $1_{10}\rightarrow 1_{01}$  & 0.0035  & D02, F07, G93 \\
                       &  180.49  &  $4_{23}\rightarrow 3_{12}$  &  0.0306  & \\
                       &  179.53  &  $2_{12}\rightarrow 1_{01}$  & 0.0559   & \\
                       &  174.63  &  $3_{03}\rightarrow 2_{12}$  & 0.0505   & \\
                       &  108.07  &  $2_{12}\rightarrow 1_{10}$  & 0.256   & \\
                       &    78.74  &  $4_{23}\rightarrow 3_{12}$  & 0.484   & \\
\hline
 p-H$_2$O   &  303.46  &  $2_{02}\rightarrow 1_{11}$  & 0.0058   & D02, F07, G93 \\
                       & 269.27 &   $1_{11}\rightarrow 0_{00}$   & 0.0184   & \\
                      &  144.52  &   $4_{13}\rightarrow 3_{22}$ &  0.0332  & \\
                      &  138.53  &    $3_{13}\rightarrow 2_{02}$ & 0.125   & \\
                      &  100.98  &   $2_{20}\rightarrow 1_{11}$  &  0.260  & \\
                      &     89.99 &   $3_{22}\rightarrow 2_{11}$  &  0.352  & \\[2mm]
\end{tabular}
{References: B98 \citep{Bell1998}, C80 \citep{Chambaud1980}, D02 \citep{Dubernet2002}, F07 \citep{Faure2007}, F77 \citep{Flower1977}, F01 \citep{Flower2001}, G93 \citep{Green1993}, J92 \citep{Jaquet1992}, J05 \citep{Jankowski2005}, L77 \citep{Launay1977}, W06 \citep{Wernli2006}, W02 \citep{Wilson2002}, Y10 \citep{Yang2010}}
\label{linelist}
\end{table}

%\subsection{Parameter space}
%\label{ParSpace}

\subsection{Disk stability}
\label{Stability}

We choose here the Toomre criterium to check our disk models against gravitational instabilities. The Toomre $Q$ parameter is defined as
\begin{equation}
Q(r) = \frac{\kappa(r) c_s(r)}{\pi G \Sigma(r)}
\label{Eq:Q}
\end{equation}
where $c_s$ is the sound speed, $\Sigma$ the surface density, $G$ the gravitational constant and the epicyclic frequency $\kappa$ is defined as
\begin{equation}
\kappa(r)^2 = \frac{1}{r^3} \frac{d(r^4 \Omega(r)^2)}{dr}
\label{Eq:kappa}
\end{equation}
with $\Omega$ the orbital frequency. The disk is unstable at a certain distance $r$ from the star if $Q(r)<1$. In the following, we derive an expression that provides a simple analytical way of checking 'global' disk instability. For that, we make a few simplifying assumptions
\begin{enumerate}
\item The disks rotate with keplerian frequency $\Omega_{\rm kep} = \sqrt{GM_\ast/r^3}$, hence $\kappa(r) = \Omega_{\rm kep}(r)$
\item The midplane gas temperature can be approximated by a radial power law $T = T_0 (r/R_{\rm in})^{-0.5}$.
\end{enumerate}
We will discuss the impact of these simplifications further below.

Inserting the power laws for temperature and surface density into Eq.(\ref{Eq:Q}), we obtain
\begin{equation}
 Q(r) = \sqrt{\frac{k T_0 M_\ast}{\mu m_H G \pi^2 R_{\rm in}^3 \Sigma_0^2}} \,\, \left(\frac{r}{R_{\rm in}}\right)^{\epsilon - 7/4}
\label{Eq:Q1}
\end{equation}
Here $T_0$ and $\Sigma_0$ denote the gas temperature and surface density at the inner radius of the disk $R_{\rm in}$. Since for our values of $\epsilon$ (0.5, 1.0, 1.5) $Q(r)$ decreases with radius, disk models that obey $Q>1$ at the outer radius are also stable at smaller radii. Hence, in the following, we derive an expression for $Q$ at  $r=R_{\rm out}$, which can be used to determine the global stability of a disk model.

The surface density and temperature at the inner radius can be derived from the total disk gas mass $M_{\rm disk}$ and the luminosity and effective temperature of the central star $L_\ast$, $T_{\rm eff}$ \citep{Tuthill2001,vanderPlas2010}
\begin{eqnarray}
\Sigma_0 & = & \frac{M_{\rm disk} \left(2-\epsilon \right)}{4 \pi R_{\rm in}^2} \left( \left(\frac{R_{\rm out}}{R_{\rm in}}\right)^{(2-\epsilon)} -1 \right)^{-1} \\
T_0 & = & \sqrt{\frac{R_\ast}{R_{\rm in}}} T_\ast 
\end{eqnarray}
Using these, we can rewrite Eq.(\ref{Eq:Q1}) entirely in terms of the input parameters of the {\sc Dent} grid
\begin{eqnarray}
 Q & = & \sqrt{\frac{8 k}{\mu m_H G }} \sqrt{ \frac{\sqrt{R_\ast} T_\ast M_\ast }{ M_{\rm disk}^2 (2-\epsilon)^2}} \,\,\, R_{\rm in}^{1/4} \\ \nonumber
 & & \left( \left(\frac{R_{\rm out}}{R_{\rm in}}\right)^{(2-\epsilon)} -1 \right) \left(\frac{R_{\rm out}}{R_{\rm in}}  \right)^{(\epsilon - 7/4)} 
\end{eqnarray}

We apply this criterium to all 322\,030 disk models and find that only 16422 of them are unstable according to the above derived criterium. Hence, 94\% of all models are not affected by gravitational instabilities. This percentage changes by less than 1\% if we (a) choose a more realistic temperature profile for optically thick disks $T \sim (r/R_{\rm in})^{-0.25}$ or (b) change $T_0$ by a factor 2. The fact that $Q(r)$ decreases with $r$ still holds for $T \sim (r/R_{\rm in})^{-0.25}$. Overall, we find that models with high gas disk masses --- as expected --- are more subject to instabilities. However, even in our highest mass bin the fraction of unstable models is only $\sim 30$\%. The plots and conclusions of the analysis of the {\sc DENT} grid are not affected by the small number of models that could become unstable.

\subsection{Methodology}
\label{Methodology}

The large number of disk models makes it impossible to carry out an individual study of their chemical and thermal structure. Hence in the following, we study the statistics of species gas temperatures and masses and how they depend on certain grid parameters such as the surface density distribution and settling.

For the statistical analysis, we define a scalar quantity $X_i$ of any model $i$. It can be a parameter, or can be a measured value like the total mass of atomic oxygen $M({\rm O})$ of model $i$. Next, we select an interval on which we perform the statistics $[X_k-dx/2, X_k+dx/2]$, where $X_k$ are sampling points for $X$. The total number of models (selected $i$) in that selected range are $N_{\rm sel}$. The expected value and standard deviation of the distribution are then defined as
\begin{eqnarray}
\langle X \rangle & = & \sum_{{\rm selected}~i} X_i /N_{\rm sel}  \\
\sigma  & = & \sqrt{ \sum_{{\rm selected}~i} (X_i - \langle X \rangle)^2 /N_{\rm sel}  } \,\,\, .
\end{eqnarray}

\bibliography{reference}

\begin{thebibliography}{61}
\expandafter\ifx\csname natexlab\endcsname\relax\def\natexlab#1{#1}\fi

\bibitem[{{Aikawa} \& {Nomura}(2006)}]{Aikawa2006}
{Aikawa}, Y. \& {Nomura}, H. 2006, \apj, 642, 1152

\bibitem[{{Aikawa} {et~al.}(2002){Aikawa}, {van Zadelhoff}, {van Dishoeck}, \&
  {Herbst}}]{Aikawa2002}
{Aikawa}, Y., {van Zadelhoff}, G.~J., {van Dishoeck}, E.~F., \& {Herbst}, E.
  2002, \aap, 386, 622

\bibitem[{{Aresu} {et~al.}(2011){Aresu}, {Kamp}, {Meijerink}, {Woitke}, {Thi},
  \& {Spaans}}]{Aresu2011}
{Aresu}, G., {Kamp}, I., {Meijerink}, R., {et~al.} 2011, \aap, 526, A163+

\bibitem[{{Bell} {et~al.}(1998){Bell}, {Berrington}, \& {Thomas}}]{Bell1998}
{Bell}, K.~L., {Berrington}, K.~A., \& {Thomas}, M.~R.~J. 1998, \mnras, 293,
  L83

\bibitem[{{Bjorkman} \& {Wood}(2001)}]{Bjorkman2001}
{Bjorkman}, J.~E. \& {Wood}, K. 2001, \apj, 554, 615

\bibitem[{{Brott} \& {Hauschildt}(2005)}]{Brott2005}
{Brott}, I. \& {Hauschildt}, P.~H. 2005, in ESA Special Publication, Vol. 576,
  The Three-Dimensional Universe with Gaia, ed. C.~{Turon}, K.~S. {O'Flaherty},
  \& M.~A.~C. {Perryman}, 565--+

\bibitem[{{Cernicharo} {et~al.}(2009){Cernicharo}, {Ceccarelli}, {M{\'e}nard},
  {Pinte}, \& {Fuente}}]{Cernicharo2009}
{Cernicharo}, J., {Ceccarelli}, C., {M{\'e}nard}, F., {Pinte}, C., \& {Fuente},
  A. 2009, \apjl, 703, L123

\bibitem[{{Chambaud} {et~al.}(1980){Chambaud}, {Levy}, {Millie}, \& {et
  al.}}]{Chambaud1980}
{Chambaud}, G., {Levy}, B., {Millie}, P., \& {et al.} 1980, Journal of Physics
  B Atomic Molecular Physics, 13, 4205

\bibitem[{{Chiang} \& {Goldreich}(1997)}]{Chiang1997}
{Chiang}, E.~I. \& {Goldreich}, P. 1997, \apj, 490, 368

\bibitem[{{D'Alessio} {et~al.}(1998){D'Alessio}, {Canto}, {Calvet}, \&
  {Lizano}}]{Dalessio1998}
{D'Alessio}, P., {Canto}, J., {Calvet}, N., \& {Lizano}, S. 1998, \apj, 500,
  411

\bibitem[{{Dent} {et~al.}(2005){Dent}, {Greaves}, \& {Coulson}}]{Dent2005}
{Dent}, W.~R.~F., {Greaves}, J.~S., \& {Coulson}, I.~M. 2005, \mnras, 359, 663

\bibitem[{{Draine}(2006)}]{Draine2006}
{Draine}, B.~T. 2006, \apj, 636, 1114

\bibitem[{{Draine} \& {Lee}(1984)}]{Draine1984}
{Draine}, B.~T. \& {Lee}, H.~M. 1984, \apj, 285, 89

\bibitem[{{Dubernet} \& {Grosjean}(2002)}]{Dubernet2002}
{Dubernet}, M. \& {Grosjean}, A. 2002, \aap, 390, 793

\bibitem[{{Dullemond} \& {Dominik}(2004)}]{Dullemond2004}
{Dullemond}, C.~P. \& {Dominik}, C. 2004, \aap, 417, 159

\bibitem[{{Dullemond} {et~al.}(2002){Dullemond}, {van Zadelhoff}, \&
  {Natta}}]{Dullemond2002}
{Dullemond}, C.~P., {van Zadelhoff}, G.~J., \& {Natta}, A. 2002, \aap, 389, 464

\bibitem[{{Faure} {et~al.}(2007){Faure}, {Crimier}, {Ceccarelli}, {Valiron},
  {Wiesenfeld}, \& {Dubernet}}]{Faure2007}
{Faure}, A., {Crimier}, N., {Ceccarelli}, C., {et~al.} 2007, \aap, 472, 1029

\bibitem[{{Field} {et~al.}(1980){Field}, {Adams}, \& {Smith}}]{Field1980}
{Field}, D., {Adams}, N.~G., \& {Smith}, D. 1980, \mnras, 192, 1

\bibitem[{{Flower}(2001)}]{Flower2001}
{Flower}, D.~R. 2001, J. Phys. B, 34, 2731

\bibitem[{{Flower} \& {Launay}(1977)}]{Flower1977}
{Flower}, D.~R. \& {Launay}, J.~M. 1977, Journal of Physics B Atomic Molecular
  Physics, 10, 3673

\bibitem[{{Glassgold} {et~al.}(2009){Glassgold}, {Meijerink}, \&
  {Najita}}]{Glassgold2009}
{Glassgold}, A.~E., {Meijerink}, R., \& {Najita}, J.~R. 2009, \apj, 701, 142

\bibitem[{{Glauser} {et~al.}(2008){Glauser}, {M{\'e}nard}, {Pinte},
  {Duch{\^e}ne}, {G{\"u}del}, {Monin}, \& {Padgett}}]{Glauser2008}
{Glauser}, A.~M., {M{\'e}nard}, F., {Pinte}, C., {et~al.} 2008, \aap, 485, 531

\bibitem[{{Goicoechea} {et~al.}(2009){Goicoechea}, {Swinyard}, \& {Spica/Safari
  Science Team}}]{Goicoechea2009}
{Goicoechea}, J.~R., {Swinyard}, B., \& {Spica/Safari Science Team}. 2009, in
  SPICA joint European/Japanese Workshop, held 6-8 July, 2009 at Oxford, United
  Kingdom. Edited by A.M. Heras, B.M. Swinyard, K.G. Isaak, and J.R.
  Goicoechea. EDP Sciences, 2009, p.02002, 2002--+

\bibitem[{{Gorti} \& {Hollenbach}(2004)}]{Gorti2004}
{Gorti}, U. \& {Hollenbach}, D. 2004, \apj, 613, 424

\bibitem[{{Gorti} \& {Hollenbach}(2008)}]{Gorti2008}
{Gorti}, U. \& {Hollenbach}, D. 2008, \apj, 683, 287

\bibitem[{{Green} {et~al.}(1993){Green}, {Maluendes}, \& {McLean}}]{Green1993}
{Green}, S., {Maluendes}, S., \& {McLean}, A.~D. 1993, \apjs, 85, 181

\bibitem[{{Henning} {et~al.}(2010){Henning}, {Semenov}, {Guilloteau}, {Dutrey},
  {Hersant}, {Wakelam}, {Chapillon}, {Launhardt}, {Pi{\'e}tu}, \&
  {Schreyer}}]{Henning2010}
{Henning}, T., {Semenov}, D., {Guilloteau}, S., {et~al.} 2010, \apj, 714, 1511

\bibitem[{{Isella} {et~al.}(2010){Isella}, {Carpenter}, \&
  {Sargent}}]{Isella2010}
{Isella}, A., {Carpenter}, J.~M., \& {Sargent}, A.~I. 2010, \apj, 714, 1746

\bibitem[{{Jankowski} \& {Szalewicz}(2005)}]{Jankowski2005}
{Jankowski}, P. \& {Szalewicz}, K. 2005, JChPh 123, 10, 104301

\bibitem[{{Jaquet} {et~al.}(1992){Jaquet}, {Staemmler}, {Smith}, \&
  {Flower}}]{Jaquet1992}
{Jaquet}, R., {Staemmler}, V., {Smith}, M.~D., \& {Flower}, D.~R. 1992, Journal
  of Physics B Atomic Molecular Physics, 25, 285

\bibitem[{{Jonkheid} {et~al.}(2007){Jonkheid}, {Dullemond}, {Hogerheijde}, \&
  {van Dishoeck}}]{Jonkheid2007}
{Jonkheid}, B., {Dullemond}, C.~P., {Hogerheijde}, M.~R., \& {van Dishoeck},
  E.~F. 2007, \aap, 463, 203

\bibitem[{{Jonkheid} {et~al.}(2004){Jonkheid}, {Faas}, {van Zadelhoff}, \& {van
  Dishoeck}}]{Jonkheid2004}
{Jonkheid}, B., {Faas}, F.~G.~A., {van Zadelhoff}, G.-J., \& {van Dishoeck},
  E.~F. 2004, \aap, 428, 511

\bibitem[{{Kamp} \& {Dullemond}(2004)}]{Kamp2004}
{Kamp}, I. \& {Dullemond}, C.~P. 2004, \apj, 615, 991

\bibitem[{{Kamp} {et~al.}(2010){Kamp}, {Tilling}, {Woitke}, {Thi}, \&
  {Hogerheijde}}]{Kamp2010}
{Kamp}, I., {Tilling}, I., {Woitke}, P., {Thi}, W., \& {Hogerheijde}, M. 2010,
  \aap, 510, A260000+

\bibitem[{{Launay} \& {Roueff}(1977)}]{Launay1977}
{Launay}, J.~M. \& {Roueff}, E. 1977, \aap, 56, 289

\bibitem[{{Liseau} {et~al.}(2006){Liseau}, {Justtanont}, \&
  {Tielens}}]{Liseau2006}
{Liseau}, R., {Justtanont}, K., \& {Tielens}, A.~G.~G.~M. 2006, \aap, 446, 561

\bibitem[{{Lorenzetti} {et~al.}(2002){Lorenzetti}, {Giannini}, {Nisini},
  {Benedettini}, {Elia}, {Campeggio}, \& {Strafella}}]{Lorenzetti2002}
{Lorenzetti}, D., {Giannini}, T., {Nisini}, B., {et~al.} 2002, \aap, 395, 637

\bibitem[{{Lucy}(1999)}]{Lucy1999}
{Lucy}, L.~B. 1999, \aap, 345, 211

\bibitem[{{Meijerink} {et~al.}(2008){Meijerink}, {Glassgold}, \&
  {Najita}}]{Meijerink2008}
{Meijerink}, R., {Glassgold}, A.~E., \& {Najita}, J.~R. 2008, \apj, 676, 518

\bibitem[{{Nomura} {et~al.}(2007){Nomura}, {Aikawa}, {Tsujimoto}, {Nakagawa},
  \& {Millar}}]{Nomura2007}
{Nomura}, H., {Aikawa}, Y., {Tsujimoto}, M., {Nakagawa}, Y., \& {Millar}, T.~J.
  2007, \apj, 661, 334

\bibitem[{{Nomura} \& {Millar}(2005)}]{Nomura2005}
{Nomura}, H. \& {Millar}, T.~J. 2005, \aap, 438, 923

\bibitem[{{Pinte} {et~al.}(2009){Pinte}, {Harries}, {Min}, {Watson},
  {Dullemond}, {Woitke}, {M{\'e}nard}, \& {Dur{\'a}n-Rojas}}]{Pinte2009}
{Pinte}, C., {Harries}, T.~J., {Min}, M., {et~al.} 2009, \aap, 498, 967

\bibitem[{{Pinte} {et~al.}(2006){Pinte}, {M{\'e}nard}, {Duch{\^e}ne}, \&
  {Bastien}}]{Pinte2006}
{Pinte}, C., {M{\'e}nard}, F., {Duch{\^e}ne}, G., \& {Bastien}, P. 2006, \aap,
  459, 797

\bibitem[{{Pinte} {et~al.}(2010){Pinte}, {Woitke}, {Menard}, {Duchene}, {Kamp},
  {Meeus}, {Mathews}, {Howard}, {Grady}, {Thi}, {Tilling}, {Augereau}, {Dent},
  {Alacid}, {Andrews}, {Ardila}, {Aresu}, {Barrado}, {Brittain}, {Ciardi},
  {Danchi}, {Eiroa}, {Fedele}, {de Gregorio-Monsalvo}, {Heras}, {Huelamo},
  {Krivov}, {Lebreton}, {Liseau}, {Martin-Zaidi}, {Mendigutia}, {Montesinos},
  {Mora}, {Morales-Calderon}, {Nomura}, {Pantin}, {Pascucci}, {Phillips},
  {Podio}, {Poelman}, {Ramsay}, {Riaz}, {Rice}, {Riviere-Marichalar},
  {Roberge}, {Sandell}, {Solano}, {Vandenbussche}, {Walker}, {Williams},
  {White}, \& {Wright}}]{Pinte2010}
{Pinte}, C., {Woitke}, P., {Menard}, F., {et~al.} 2010, ArXiv e-prints

\bibitem[{{Pontoppidan} {et~al.}(2010){Pontoppidan}, {Salyk}, {Blake},
  {Meijerink}, {Carr}, \& {Najita}}]{Pontoppidan2010}
{Pontoppidan}, K.~M., {Salyk}, C., {Blake}, G.~A., {et~al.} 2010, \apj, 720,
  887

\bibitem[{{Qi} {et~al.}(2003){Qi}, {Kessler}, {Koerner}, {Sargent}, \&
  {Blake}}]{Qi2003}
{Qi}, C., {Kessler}, J.~E., {Koerner}, D.~W., {Sargent}, A.~I., \& {Blake},
  G.~A. 2003, \apj, 597, 986

\bibitem[{{Qi} {et~al.}(2008){Qi}, {Wilner}, {Aikawa}, {Blake}, \&
  {Hogerheijde}}]{Qi2008}
{Qi}, C., {Wilner}, D.~J., {Aikawa}, Y., {Blake}, G.~A., \& {Hogerheijde},
  M.~R. 2008, \apj, 681, 1396

\bibitem[{{Ricci} {et~al.}(2010){Ricci}, {Testi}, {Natta}, \&
  {Brooks}}]{Ricci2010}
{Ricci}, L., {Testi}, L., {Natta}, A., \& {Brooks}, K.~J. 2010, \aap, 521, A66+

\bibitem[{{Sch{\"o}ier} {et~al.}(2005){Sch{\"o}ier}, {van der Tak}, {van
  Dishoeck}, \& {Black}}]{Schoier2005}
{Sch{\"o}ier}, F.~L., {van der Tak}, F.~F.~S., {van Dishoeck}, E.~F., \&
  {Black}, J.~H. 2005, \aap, 432, 369

\bibitem[{{Semenov} {et~al.}(2005){Semenov}, {Pavlyuchenkov}, {Schreyer},
  {Henning}, {Dullemond}, \& {Bacmann}}]{Semenov2005}
{Semenov}, D., {Pavlyuchenkov}, Y., {Schreyer}, K., {et~al.} 2005, \apj, 621,
  853

\bibitem[{{Siess} {et~al.}(2000){Siess}, {Dufour}, \& {Forestini}}]{Siess2000}
{Siess}, L., {Dufour}, E., \& {Forestini}, M. 2000, \aap, 358, 593

\bibitem[{{Thi} {et~al.}(2004){Thi}, {van Zadelhoff}, \& {van
  Dishoeck}}]{Thi2004}
{Thi}, W.-F., {van Zadelhoff}, G.-J., \& {van Dishoeck}, E.~F. 2004, \aap, 425,
  955

\bibitem[{{Tielens} \& {Hollenbach}(1985)}]{Tielens1985}
{Tielens}, A.~G.~G.~M. \& {Hollenbach}, D. 1985, \apj, 291, 747

\bibitem[{{Tuthill} {et~al.}(2001){Tuthill}, {Monnier}, \&
  {Danchi}}]{Tuthill2001}
{Tuthill}, P.~G., {Monnier}, J.~D., \& {Danchi}, W.~C. 2001, Nature, 409, 1012

\bibitem[{{van der Plas} {et~al.}(2011){van der Plas}, {van den Ancker},
  {Waters}, \& {Dominik}}]{vanderPlas2010}
{van der Plas}, G., {van den Ancker}, M.~E., {Waters}, L.~B.~F.~M., \&
  {Dominik}, C. 2011, A\&A, 0, submitted

\bibitem[{{Wernli} {et~al.}(2006){Wernli}, {Valiron}, {Faure}, {Wiesenfeld},
  {Jankowski}, \& {Szalewicz}}]{Wernli2006}
{Wernli}, M., {Valiron}, P., {Faure}, A., {et~al.} 2006, \aap, 446, 367

\bibitem[{{Wilson} \& {Bell}(2002)}]{Wilson2002}
{Wilson}, N.~J. \& {Bell}, K.~L. 2002, \mnras, 337, 1027

\bibitem[{{Woitke} {et~al.}(2009{\natexlab{a}}){Woitke}, {Kamp}, \&
  {Thi}}]{Woitke2009a}
{Woitke}, P., {Kamp}, I., \& {Thi}, W.-F. 2009{\natexlab{a}}, \aap, 501, 383

\bibitem[{{Woitke} {et~al.}(2010){Woitke}, {Pinte}, {Tilling}, {M{\'e}nard},
  {Kamp}, {Thi}, {Duch{\^e}ne}, \& {Augereau}}]{Woitke2010}
{Woitke}, P., {Pinte}, C., {Tilling}, I., {et~al.} 2010, \mnras, L53+

\bibitem[{{Woitke} {et~al.}(2009{\natexlab{b}}){Woitke}, {Thi}, {Kamp}, \&
  {Hogerheijde}}]{Woitke2009b}
{Woitke}, P., {Thi}, W.-F., {Kamp}, I., \& {Hogerheijde}, M.~R.
  2009{\natexlab{b}}, \aap, 501, L5

\bibitem[{{Yang} {et~al.}(2010){Yang}, {Stancil}, {Balakrishnan}, \&
  {Forrey}}]{Yang2010}
{Yang}, B., {Stancil}, P.~C., {Balakrishnan}, N., \& {Forrey}, R.~C. 2010,
  \apj, 718, 1062

\end{thebibliography}

%\begin{thebibliography}{}
%\end{thebibliography}

\end{document}